\documentclass[reviewcopy]{elsart}

\usepackage{natbib}
\usepackage{lineno}

\usepackage{epsfig}

\usepackage{amssymb}

\newcommand{\methane}{CH$_4$}

\newcommand{\ammonia}{NH$_3$}
\newcommand{\phosphine}{PH$_3$}

\newcommand{\hydrogen}{H$_{2}$}

\newcommand{\acet}{C$_{2}$H$_{2}$}

\newcommand{\ethane}{C$_{2}$H$_{6}$}

\newcommand{\micron}{$\mu$m}
\newcommand{\microbar}{$\mu$bar}
\newcommand{\cm}{cm$^{-1}$}
\newcommand{\radunit}{${\rm W \: cm^{-2} \: sr^{-1} / cm^{-1}}$}

\newcommand{\dg}{$^{\circ}$}

\def\baselinestretch{1.6}

\begin{document}
\linenumbers

\parindent 0mm
\Large

\centerline{\bf Abundances of Jupiter's Trace Hydrocarbons}
\centerline{\bf From Voyager and Cassini}

\vspace*{1cm}
\large 
\centerline{C.A.~Nixon$_{a,b}$, R.K. Achterberg$_{a,b}$,
  P.N. Romani$_{b}$, M.~Allen$_{c,d}$,}
\centerline{X.~Zhang$_c$, N.A.~Teanby$_e$, P.G.J.~Irwin$_e$, F.M.~Flasar$_b$} 

\vspace*{0.5cm}
\centerline{$^*$ Corresponding author: {\tt e-mail: conor.a.nixon@nasa.gov}}

\vspace*{0.5cm}
\normalsize
\centerline{$_a$University of Maryland, College Park, MD 20742, USA}
\centerline{$_b$NASA Goddard Space Flight Center, Greenbelt, MD 20771,
USA}
\centerline{$_c$California Institute of Technology, Pasadena, CA 91125}
\centerline{$_d$Jet Propulsion Laboratory, 4800 Oak Grove Drive, 
Pasadena, CA 91109}
\centerline{$_e$Atmospheric, Oceanic and Planetary Physics,
University of Oxford, Clarendon Laboratory,}
\centerline{Parks Road, Oxford, OX1 3PU, UK}
\large
\vspace*{1cm}
\centerline{\bf Accepted for publication in {\em Planetary and Space Science}, May 17 2010}
\vspace*{0.5cm}
\centerline{doi 10.1016/j.pss2010.05.08}
%\centerline{For submission to {\it  Planetary and Space Science} }
\vspace*{0.5cm}
\centerline{(figures=12 tables=5)}
\newpage

%========SECOND PAGE

\Large
Running Head: \newline
\centerline{\bf Jupiter's Hydrocarbon Abundances}

\vspace*{3cm}
\large
Direct correspondence to: \newline
Conor A. Nixon \newline
Solar System Exploration Division \\
Planetary Systems Laboratory - Code 693 \\
NASA Goddard Space Flight Center \newline
Greenbelt \newline
MD 20771 \newline
U.S.A. \newline
tel. (301) 286-6757 \newline
fax. (301) 286-0212 \newline

\newpage
%%%%%%%%%%%%%%%%%%%%%%%%%%%%%%%%%%%%%%%%%%%%%%%%%%%%%%%%%%%%%%%%%%%%%%%%%
%%%%%%%%%%%%%%%%%%%%%%%%%%%%%%%%%%%%%%%%%%%%%%%%%%%%%%%%%%%%%%%%%%%%%%%%%
%========ABSTRACT

\parindent 6mm
\LARGE
\centerline{\bf ABSTRACT}
\normalsize 

The flybys of Jupiter by the Voyager spacecraft in 1979, and over two
decades later by Cassini in 2000, have provided us with unique
datasets from two different epochs, allowing the investigation of
seasonal change in the atmosphere. In this paper we model zonal
averages of thermal infrared spectra from the two instruments, Voyager
1 IRIS and Cassini CIRS, to retrieve the vertical and meridional
profiles of temperature, and the abundances of the two minor
hydrocarbons, acetylene (\acet ) and ethane (\ethane ). The spatial
variation of these gases is controlled by both
chemistry and dynamics, and therefore their observed distribution
gives us an insight into both processes. We find that the two gases
paint quite different pictures of seasonal change. Whilst the 2-D
cross-section of \ethane\ abundance 
is slightly increased and more symmetric in 2000
(northern summer solstice) compared to 1979 (northern fall equinox),
the major trend of equator to pole increase remains. For \acet\ on the
other hand, the Voyager epoch exhibits almost no latitudinal variation,
whilst the Cassini era shows a marked decrease polewards in both
hemispheres. At the present time, these experimental findings are in
advance of interpretation, as there are no published models of 2-D Jovian
seasonal chemical variation available for comparison.

\vspace*{1cm}
\centerline{\bf key words=JUPITER ATMOSPHERE; ATMOSPHERIC ABUNDANCES,}
\centerline{\bf OUTER PLANETS; INFRARED SPECTROSCOPY; ABUNDANCE RETRIEVAL}

\newpage
\normalsize
%========MAIN TEXT

%%%%%%%%%%%%%%%%%%%%%%%%%%%%%%%%%%%%%%%%%%%%%%%%%%%%%%%%%%%%%%%%%%%%%%%%%
%%%%%%%%%%%%%%%%%%%%%%%%%%%%%%%%%%%%%%%%%%%%%%%%%%%%%%%%%%%%%%%%%%%%%%%%%
\section{Introduction}
\label{intro}

The Voyager 1 and 2 encounters with Jupiter of March and July 1979 
(jovian northern fall equinox) offered
the first opportunities to map Jupiter in the thermal
infrared with a latitude resolution of better than 10\dg\ - then
unobtainable from the ground - and a spectral resolution sufficient to
measure the abundances of trace gas species. The tool for this mapping
was IRIS \citep[the InfraRed Interferometer and
  Spectrometer,][]{hanel77}, and trace gases of interest included
ethane (\ethane ) and acetylene (\acet ) - secondary species derived
from photolysis of the primary carbon-bearing molecule, methane
(\methane ). Although some preliminary findings were published in a
conference report \citep{maguire84}, a full radiative transfer model
was never applied to retrieve a meridional trend in the abundances.

Two decades later, the Cassini spacecraft swung by Jupiter in a
distant flyby maneuver (137 R$_{\rm J}$) en route to Saturn. The
official encounter period, during which time
observations were made, lasted six months 
symmetric about the closest approach on
December 30th 2000 (northern summer solstice). Carrying
on board the Composite Infrared Spectrometer (CIRS), Cassini was able
to re-map the planet in the thermal infrared, achieving a 
spatial resolution comparable to Voyager IRIS
but with much higher maximum spectral resolution (0.48 \cm ,
versus 3.9 \cm\ for IRIS, full width to half maximum (FWHM)). 
Meridional trends of stratospheric and tropospheric
\acet\ and \ethane\ have been retrieved from CIRS data acquired during
December 1-15 2000 prior to cloest approach, by \citet{nixon07}
(hereafter Paper I).

The abundances of the minor hydrocarbons are important because these species
serve as tracers of atmospheric circulation. The photochemical 
lifetime of ethane in
the stratosphere (3$\times 10^{10}$~s at 5 mbar) is significantly
greater than one Jovian year (3.7$\times 10^{8}$~s), whereas the
acetylene lifetime (5$\times 10^{7}$~s at 5 mbar) is substantially
shorter. Note that the solar cycle (3.4$\times 10^{8}$~s) which
affects the photochemistry almost equals the Jovian year.
In the presence of dynamical motions on seasonal timescales,
the meridional abundance trend in a relatively short-lived species
(such as acetylene) is expected to show significant change, 
whereas a long-lived species (ethane) should present a more constant
distribution. The logical means of testing
this hypothesis is to intercompare the hydrocarbon abundances using
the datasets of Voyager and Cassini, which encountered Jupiter at two
different seasons: near northern fall equinox, and 1.75 jovian years
later just after northern summer solstice. 

Such a comparison is the objective of this paper. 
We begin by analyzing the Voyager spectra, retrieving for
the first time with a full radiative transfer modeling approach the
meridional abundance trends of ethane and acetylene in 1979. We also
reanalyze the CIRS dataset, using a significantly revised
spectroscopic atlas for ethane \citep{auwera07} that was published
since the first Cassini jovian hydrocarbons paper appeared (and is
also used the IRIS data analysis here). The intensities of this new
list are revised upwards by some 44\% relative to the GEISA 2003 list
used in Paper I, resulting in a corresponding decrease in
abundances. We then proceed to compare the abundance profiles at the
two epochs (northern fall in 1979 versus summer in 2000) discussing
the implications and drawing conclusions. 

%%%%%%%%%%%%%%%%%%%%%%%%%%%%%%%%%%%%%%%%%%%%%%%%%%%%%%%%%%%%%%%%%%%%%%%%%
%%%%%%%%%%%%%%%%%%%%%%%%%%%%%%%%%%%%%%%%%%%%%%%%%%%%%%%%%%%%%%%%%%%%%%%%%
\section{Data Acquisition}

\subsection{Cassini CIRS Observations}
\label{sect:cirsobs}

The Casini CIRS instrument is a dual interferometer, with a mid-IR Michelson
interferometer covering the range 600--1400~\cm , and a far-IR
polarizing interferometer covering the range 10--600~\cm\
\citep{flasar04b}. The spectra analyzed here have been selected in the
same way as in Paper I (described below), the only difference being that they
are derived from a later version of the CIRS database. This has some
improvements to the calibration algorithm, including rejection of some
`bad' spectra (scan mechanism out of phase 
lock) and greater suppression of numerical
artifacts from the Fourier transform. These mid-IR spectra were
acquired at the highest spectral resolution of CIRS (0.48~\cm\ FWHM,
Hamming apodized) during December 1--15 2000, and cover a latitude
range of 70\dg S to 70\dg N with a spatial resolution of 6.0--3.5\dg\
great circle arc. The spectra were binned in 5\dg\ bins, and with
auroral areas excluded as before. The effective emission angle
$\bar\theta$ for the $i^{th}$ bin was computed from the mean of
the airmass (plane parallel atmosphere) of the $N_j$ spectra as follows: 

\begin{equation}
\bar\theta_i = {\rm arcsec}  \left[ \frac{1}{N_j} \sum_{j=1}^{N_j} \sec \left(
\theta_j \right) \right]
\end{equation}

\noindent 
For further details see Table I. 

\centerline{\bf [TABLE I appears here]}

\subsection{Voyager IRIS Observations}
\label{sect:irisobs}

The two Voyager spacecraft passed much closer to Jupiter than Cassini
($5.7\times10^{5}$~km versus $9.8\times10^{6}$~km). The IRIS
spectrometer carried on board was also a Michelson type - like the
mid-infrared of CIRS - but with a single bolometer detector instead of
two $1\times10$ arrays. The IRIS spectra selected here are from the
Voyager 1 inbound north/south map, taken over a 10.5 hour
period beginning 2 days, 13 hours before closest approach at a range between
$2.9\times10^{6}$ and $3.6\times10^{6}$ km. Nevertheless, the much larger
field of view (FOV) of the IRIS bolometer (4.36 mrad) versus the CIRS
mid-IR detectors \citep[0.29 mrad][]{nixon09a} results in slightly
larger projected field of view on the disk of 10 to 12 degrees of
latitude at disk center. The observation was made by repeatedly
stepping the field of view from north to south along Jupiter's central
meridian, with the rotation of Jupiter providing longitude
coverage. The spectra were then averaged in 10\dg\ wide latitude bins, 
stepped every 5\dg . As the spectra were acquired at constant airmass
for each latitude, the emission angles were simply averaged in each bin.
See Table II for details.

\centerline{\bf [TABLE II appears here]}

%%%%%%%%%%%%%%%%%%%%%%%%%%%%%%%%%%%%%%%%%%%%%%%%%%%%%%%%%%%%%%%%%%%%%%%%%
%%%%%%%%%%%%%%%%%%%%%%%%%%%%%%%%%%%%%%%%%%%%%%%%%%%%%%%%%%%%%%%%%%%%%%%%%
\section{Data Analysis}
\label{method}

%%%%%%%%%%%%%%%%%%%%%%%%%%%%%%%%%%%%%%%%%%%%%%%%%%%%%%%%%%%%%%%%%%%%%%%%%
\subsection{Model Atmosphere}
\label{sect:atm}

The model atmosphere used by us is identical to that of Paper I. The
initial temperature profile was taken from the Galileo probe ASI
measurements \citep{seiff98}, from 4 bar to 0.4$\times 10^{-6}$
bar. Hydrogen and helium have uniformly mixed vertical abundances (0.863 and
0.134 respectively) \citep{niemann98, vonzahn98}. Initial vertical
abundance profiles for the three hydrocarbons (\methane , \ethane\ and
\acet ) were computed using a photochemical model, described in Paper
I. Solar maximum conditions were used for the UV flux in this
calculation for both the Cassini and Voyager epochs, as both
encounters occured at similar points near the maximum in the solar cycle.

The vertical profiles for the known constituents 
ammonia and phosphine were controlled by
a three-parameter model: (1) deep abundance; (2) `knee' pressure
level, the transition level from deep, fixed, to varying abundance;
(3) a fractional scale height used to decrease the abundance above the
`knee' level. The retrieval of ammonia and phosphine parameters from
the mid-infrared spectrum has been discussed extensively elsewhere
\citep{irwin04, abbas04, fouchet04a, achterberg06}. As these gases are
not our focus here, we will remark only briefly on them concerning
their effect on the hydrocarbon retrievals.

With regard to 
\phosphine , the spectral effect of the $\nu_4$ band (1121~\cm ) is
significant in the region 900-1200~\cm\ (see Fig. 1 of
\citet{irwin04}) and insignificant outside, including the hydrocarbon
spectral ranges considered here. Therefore its parameters were fixed
at the values ($6\times 10^{-7}$, 1.0 bar, 0.3) recommended by
\citet{irwin04}. Regarding \ammonia , the spectral effect of the
$\nu_2$ band at 950~\cm\ is most significant from $\sim$800-1000~\cm\
(although the effects range from 750-1200~\cm ) , and so primarily
affects the ethane spectral range. As it varies with latitude, it must
be included in the retrieval. For our purposes the knee pressure was
fixed at 0.7 bar and the both the deep pressure and fractional scale
height were included as free parameters, with a priori values of $(2.2
\pm 0.22) \times 10^{-4}$ and $0.15 \pm 0.05$ respectively. Note also
that ammonia is entirely separable from ethane due to the very
different spectral signature. For discussion of the latitude and
longitude variation of these retrieved parameters, see the
above-referenced papers. 

Fig. \ref{fig:gas_apriori} depicts the model atmosphere, which is
divided into 71 vertical layers.

\begin{figure}[h]
\caption{Appears Here.}
\label{fig:gas_apriori}
\end{figure}

%%%%%%%%%%%%%%%%%%%%%%%%%%%%%%%%%%%%%%%%%%%%%%%%%%%%%%%%%%%%%%%%%%%%%%%%%
\subsection{Forward Spectral Model and Retrieval Algorithm}
\label{sect:fm}

The computation of the spectrum from the model atmosphere was carried
out using the {\tt Nemesis} computer code \citep{irwin08}, which also
accomplishes the fitting and parameter retrieval. 
This code has been significantly validated
in the past, having been successfully applied to model reflection and
thermal infrared spectra of Venus, Mars, Jupiter, Saturn and Uranus;
from a variety of spacecraft and ground-based facilities and
instruments, including Galileo, Cassini, Mars Express, Venus Express,
UIST. 

The Jovian model opacity is derived from two sources: (i) gas
vibrational-rotational bands, and (ii) collisional-induced gas
opacity. Jupiter's atmosphere is also known to contain stratospheric
haze. The addition of haze (particulate) opacity was investigated in
paper I using the refractive index co-efficients of \citet{khare84} or a
simple grey absorber. However, the effect on the retrieval was not
found to be statistically significant, i.e. the difference to the $\chi^2$ was
below the noise threshold. Therefore we have followed the approach of
Paper I and do not include a haze opacity in our retrievals.
The atlases for the gas bands are identical to Paper I, except for
ethane, where we substituted the more recent, improved list of
\citet{auwera07}, now also available in HITRAN 2008 \citep{rothman08}
and the 2009 update of GEISA \citep{husson08}.  The collision-induced
opacity uses the method of Borysow and co-workers \citep{borysow85,
  borysow88}.

The spectral calculation proceeds based on the correlated-$k$
approximation \citep{goody89a}, whereby tables of $k$-coefficients are
pretabulated in advance, at an appropriate range of pressure and
temperature grid points covering the entire atmospheric range of
interest. The spectral resolution was convolved into the $k$-tables
during the computation, so that separate tables were built for each
gas at both the IRIS and CIRS resolutions: 3.9 and 0.48 \cm\ FWHM of
Hamming function instrument line shape (ILS) respectively
\citep[see][section 2.2 for further details of the
  apodization]{nixon09b}. This constitutes a change and improvement
over the spectral modeling of Paper I, where the ILS was approximated
by a triangle function, and is one source of the slight differences
seen in the retrieved abundance of \acet\ from the Cassini CIRS
spectra. See further discussion in \S \ref{sect:compp1}.

The fitting of the spectrum to the data was achieved through the
formalism of \citet{rodgers00}, known as non-linear least squares
optimal estimation. In essence, this is similar to minimizing the
reduced $\chi^2$ difference between the model and data. However the
goodness-of-fit measure, known as the {\em cost function} $\phi$,
additionally includes a constraint (or smoothing term) from {\em a
  priori} (assumed) information about the likely range of values that
the parameters can take. The cost function is therefore the quadratic 
sum of the spectral $\chi^2$, 
plus a similar term that measures the deviation of
the solution from the {\em a priori} constraint (see \S4.2 of Paper
I). $\phi$ is minimized using the Levenburg-Marquart technique
\citep{press92}, along downhill gradients computed by
taking the numerical partial derivatives of the solution vector
(computed spectrum) with respect to the free parameters - also known
as the functional derivatives. The retrieval terminates when a pre-set
convergence limit is reached, in this case when the change in the cost
function is less than 0.2\% between iterations. For further details
regarding the {\tt Nemesis} algorithm, see \citet{irwin08}.

%%%%%%%%%%%%%%%%%%%%%%%%%%%%%%%%%%%%%%%%%%%%%%%%%%%%%%%%%%%%%%%%%%%%%%%%%
\subsection{Vertical sensitivity from functional derivatives}
\label{sect:fd}

Before commencing on the retrievals, we first used the forward model
to investigate the vertical regions of information, by computing
the functional derivatives $\delta I/\delta x_{i,j}$ of the radiance
$I$ for each of the $x_i$ parameters at each level $j$ in the model
atmosphere. The functional derivatives for the 0.48~\cm\ resolution
Cassini CIRS spectra are described in Paper I, so we here describe the
corresponding computations for the lower resolution (3.9~\cm ) Voyager
IRIS spectra, noting differences from the CIRS sensitivities.

\begin{figure}[h] 
\caption{Appears Here.}
\label{fig:temp_fds}
\end{figure}

Figs. \ref{fig:temp_fds} shows the functional derivatives for
temperature in two spectral regions: in the hydrogen S(1) continuum at
600--650~\cm\ sensitive to the upper troposphere ($\sim$200 mbar), 
and in the P and Q branches of the methane $\nu_4$ band at 1260--1310
\cm . The latter are sensitive to several altitude regions: 1--15~mbar
in the middle stratosphere, but also at $\sim$7~\microbar\ in the upper
stratosphere due to the optically thick center of the Q-branch at
1305~\cm . Therefore, by fixing the abundances of \hydrogen\ and
\methane , we can sound the temperature structure at three different
vertical levels.

Figs. \ref{fig:c2h2_fds}--\ref{fig:c2h6_fds} show the
functional derivatives for the gases \acet\ and \ethane . Acetylene
has a broad sensitivity of about 0.1--7~mbar in the stratosphere, but
also has absorption at 300--800~mbar in the troposphere. Similarly,
ethane has a range of 1--10~mbar the stratosphere, and 100--700~mbar
in the troposphere. 

\begin{figure}[h]
\caption{Appears Here.}
\label{fig:c2h2_fds}
\end{figure}

\begin{figure}[h]
\caption{Appears Here.}
\label{fig:c2h6_fds}
\end{figure}

Fig. \ref{fig:comp_fds_equat} summarizes the vertical sensitivity ranges of
Voyager IRIS, and compares these values to those of Cassini CIRS, both
calculated at the equator. See
also Table III. The most important difference is that the hotband of
\acet\ at $\sim$731~\cm , about 1~\cm\ in width, is resolved at the CIRS
but not at the IRIS spectral resolution. Therefore, the elevated range
of sensing for CIRS at 731~\cm\ (peak at 0.1 mbar, FWHM extending to
0.01 mbar) relative to the $\nu_5$ Q-branch at 729~\cm\ (peak at 5
mbar, FWHM to 0.03 mbar) is lost. Fig. \ref{fig:comp_fds_north} and
Table IV show similar information, but calculated at 58\dg N showing
that the sensitivity ranges shift slightly with latitude.

\centerline{\bf [TABLE III appears here]}

\begin{figure}[h]
\caption{Appears Here.}
\label{fig:comp_fds_equat}
\end{figure}

\centerline{\bf [TABLE IV appears here]}

\begin{figure}[h]
\caption{Appears Here.}
\label{fig:comp_fds_north}
\end{figure}

Before leaving this topic, we also must consider whether the vertical
retrieval range will be limited by the transition from local
thermodynamic equilibrium (LTE) to non-LTE (NLTE) conditions. For the
case of Jupiter's methane at least, the problem has been previously
considered by \citet{halthore94}, who investigated the relaxation of
dilute gaseous methane in bulk hydrogen. In that paper, the authors used a
room temperature measurement for the collisional de-excitation time
$\tau_{\rm C} = C_{10}^{-1}$ for
the $\nu_3$ band of methane (3.3 \micron ) in combination with a
Landau-Teller formula to extrapolate the relaxation to the ground
state under Jovian conditions. The same $\tau_{\rm C}$ also applies to the
$\nu_4$ band, as there is fast transfer of vibrational energy (V-V)
between these excitations. Their Equation 5 may be re-arranged:

\begin{equation}
\tau_{\rm C} = 2.3175\times 10^{-7} p^{-1} \exp (40 T^{-1/3})
\end{equation}

\noindent
where $p$ is pressure in Torr and $T$ is temperature in K. In
Fig. \ref{fig:nlte} we compare the altitude-variation of $\tau_{\rm
  C}$ calculated for our initial equatorial atmosphere, versus the
spontaneous emission timescale $\tau_{\rm A} = A_{10}^{-1}$, taken as
0.4~s following \citet{halthore94}. 
We find that $\tau_{\rm A} \simeq \tau_{\rm C}$ at a pressure of $\sim
  0.8$~\microbar . We have also made an independent estimate of
  $A_{10}$ by averaging across the \methane\ $\nu_4$ band in HITRAN
  from 1215--1315~\cm , arriving at an even longer timescale of 1.7~s,
  putting the transition level well above the upper pressure limit our
  model atmosphere.
The variation with latitude is small (using the
  profiles derived later in this paper). Therefore we conclude that
  our temperature retrievals are valid over the range indicated by the
  contribution functions, which probe pressures $p > 0.8$~\microbar .

\begin{figure}[h]
\caption{Appears Here.}
\label{fig:nlte}
\end{figure}

%%%%%%%%%%%%%%%%%%%%%%%%%%%%%%%%%%%%%%%%%%%%%%%%%%%%%%%%%%%%%%%%%%%%%%%%%
%%%%%%%%%%%%%%%%%%%%%%%%%%%%%%%%%%%%%%%%%%%%%%%%%%%%%%%%%%%%%%%%%%%%%%%%%
\section{Results}
\label{sect:results}

The retrieval proceeded in the following two-step manner. Firstly, each
latitudinal average spectrum was analyzed to retrieve temperatures in
the troposphere and stratosphere, using spectral portions of the
hydrogen continuum (600--670~\cm\ and 760--800~\cm ) and methane 
$\nu_4$ band (1225--1325~\cm ). Secondly, the temperature
profile was fixed and the gas abundances of \acet\ and \ethane\
allowed to vary, and their vertical distributions retrieved from their
bands at 670--760~\cm\ (\acet\ $\nu_5$) and 800--850~\cm\ (\ethane\
$\nu_9$).

Note that for reference, the retrieved 2-D temperature fields, gas
mixing ratios for \acet\ and \ethane , and formal errors on these
quantities, from both datasets (Voyager and Cassini epochs), have been
archived on-line. See {\tt http://hdl.handle.net/2014/41479} and {\tt
http://hdl.handle.net/2014/41480}. For this reason, extensive tables
of numeric quantities and errors are not enumerated in this paper,
while some selected values are discussed in context.

\subsection{Temperatures}
\label{sect:temp_results}

\begin{figure}[h]
\caption{Appears Here.}
\label{fig:temp_contours}
\end{figure}

Fig. \ref{fig:temp_contours} shows the retrieved temperatures for
Voyager IRIS (top), Cassini CIRS (middle) and the difference
(bottom). Considering first the  stratosphere at 1~mbar, we see
that the northern hemisphere has warmed considerably ($\sim$5~K) at
this epoch, equivalent to 1.75 Jovian years later. Overall, the
Cassini lower stratosphere exhibits a more pronounced hemispheric
asymmetry than the Voyager era, however the upper stratosphere
(0.01~mbar) is colder ($\sim -10$~K). Evidence for seasonal change
is less apparent in the troposphere at 100-400~mbar, as expected due
to the much greater thermal inertia at these levels.

Note that the upper stratosphere region (0.01~mbar) was not considered
in the earlier analysis of \citet{simon-miller06}, who also compare
retrieved Jovian temperatures from Voyager and Cassini with an
emphasis on derivation of wind fields. These authors used a
different set of Cassini maps (from early January 2001) at
lower (2.8~\cm ) spectral resolution than our selection (0.48~\cm ), 
although with higher spatial sampling in latitude and longitude. Where
our results overlap those of \citet{simon-miller06}, the agreement is
good (see their Fig. 2).

We also see evidence for the the so-called jovian `Quasi-Quadrennial
Oscillation' (QQO), a cyclic temperature variation, most prominent in
the lower stratosphere, that appears to observe an approximately
four (terrestrial) year period, and may be analogous to the Earth's
quasi-biennial oscillation (QBO, $\sim$26~month period). See
Fig. \ref{fig:temp_contours} (c), at 20~mbar and 10\dg S, where a
localized increase in temperature of $\sim$8~K is apparent. The
four-year periodicity cannot be inferred from these two widely-spaced
epochs alone, but when the time series of equatorial stratospheric
temperatures is plotted with higher temporal resolution, e.g. Fig.~4 of
\citet{simon-miller06}, the pattern becomes apparent. The four-year
cycle is not exact (hence `quasi'): periodogram analysis by
\citet{simon-miller06} indicates two peaks with periodicities of 3.5
and 4.3 years. \citet{leovy91} have suggested that the QQO is caused
by the stresses induced by vertically-propagating waves. For further
details and discussion see \citet{leovy91,friedson99b,moses04,simon-miller06}. 

\subsection{Gas abundances}
\label{sect:gas_results}

The corresponding contour maps for the hydrocarbon abundances are
shown in Figs. \ref{fig:c2h2_contours} and
\ref{fig:c2h6_contours}. Note that the vertical range of \acet\
derived from CIRS data has been cut to match the smaller vertical
range of IRIS (see \S \ref{sect:fd}).

\begin{figure}[h]
\caption{Appears Here.}
\label{fig:c2h2_contours}
\end{figure}

The \acet\ difference plot (Fig.~\ref{fig:c2h2_contours} (c)) shows the
most dramatic difference, with the polar stratospheric abundances
dropping at 1--10~mbar, and more intensely in the south than
the north. This is also evident in Fig.~\ref{fig:c2h2_contours} (b),
where the upward-curved contours contrast markedly with the more level
contours of Fig.~\ref{fig:c2h2_contours} (a).

\begin{figure}[h]
\caption{Appears Here.}
\label{fig:c2h6_contours}
\end{figure}

Ethane, in contrast, shows a much smaller change from Voyager to
Cassini, although the Cassini era appears to have a more uniform
stratospheric distribution, whereas the Voyager period is
characterized by a positive bias towards the north. Also, the CIRS
epoch shows systematically higher abundances over much of the atmosphere.

%%%%%%%%%%%%%%%%%%%%%%%%%%%%%%%%%%%%%%%%%%%%%%%%%%%%%%%%%%%%%%%%%%%%%%%%%
\section{Discussion}
\label{sect:discussion}

\subsection{Comparison to Paper I}
\label{sect:compp1}

\begin{figure}[h]
\caption{Appears Here.}
\label{fig:cirs_comp}
\end{figure}

The first reference point for this study is the previously published
retrievals of hydrocarbon abundances from CIRS data (Paper I). The
deviation between the present work and Paper I is plotted in
Fig. \ref{fig:cirs_comp}.
For \acet , the results are almost identical in regions where the
information content is maximum, e.g. at 2 mbar in the equatorial
stratosphere the change is just 1\%, and at 300 mbar in the
equatorial troposphere the change is 10\%. The maximum change was at
62\dg S, 300 mbar where the deviation reached 25\%. These differences 
are well within the overall error bars of $\sim$40-50\%, and may be
attributed to small changes in the modeling method, such as the
improved instrumental line shape (see \S \ref{sect:fm}). See on-line
data for full data and errors (\S \ref{sect:results}).

For \ethane , the change at 5 mbar in the equatorial
stratosphere was 40\%, with a 20\% change at 300 mbar. However, these
differences are relative to smaller error bars of 15-20\%, and are 
therefore clearly significant. These changes can be explained by
the new line list used for \ethane\ (see \S \ref{sect:fm}), which
has higher line intensities than used in Paper I, by some 44\% over
the whole band.

\subsection{Comparison to other previous studies}
\label{sect:comp}

In the recent review on the Jovian stratosphere by 
\citet[][\S 7.2.3 ]{moses04}, describes the history of abundance
measurements of \acet\ and \ethane\ in the Jovian non-auroral stratosphere, 
spanning the period from 1973 to 1997. Their Table 7.1 lists all major
observational
measurements, totaling 19 published studies in all, broken down into
16 of \acet\ and 23 of \ethane . These varied researches cover Jupiter's
stratosphere, mainly at low latitudes, but encompassing every season,
and at pressures from 80 to 0.005 mbar.

In light of the present findings, it is now clear that Jupiter's
stratosphere is a dynamic environment that exhibits seasonal change
in the abundances of the trace hydrocarbons at all latitudes and
pressure levels. Therefore, direct comparison between our findings and
these previous studies is not universally appropriate. Clearly, the most
meaningful way to proceed is to compare results obtained at similar
seasons to the results we have obtained for the Cassini and Voyager epochs.
Table V compares a relevant subset of observations, taken from
\citet[][Table 7.1]{moses04}, that apply to the same season as either the
Voyager or Cassini encounters. We have added error bars where those
were available in the original sources. 

\centerline{\bf [TABLE V appears here]}

Relative to our Cassini results,
we found only one applicable comparison, for \ethane\ only, which is 
the study of \citet{livengood93} based on data from 12/89, 
about 1 full Jovian year before Cassini. The Cassini-derived value of
$(9.4\pm 1.8)$~ppm at 1~mbar is somewhat greater their value of
$(3.6\pm 1.3)$ ppm, at 0.3--3 mbar, similar to our conclusion in Paper
I. We note however that we would achieve agreement at
their lower bound of 3~mbar where we find an ethane VMR of 
$(4.8\pm 0.8)$~ppm. The other studies for \ethane\ in 
the 4-5 mbar region also agree with our results when the
respective error bars are considered. Note that the entire set of past
measurements for \ethane\ is generally self-consistent, in marked
contrast to those of \acet , discussed hereafter.

With regard to \acet , our results are compatible with those of
\citet{owen80} (10 mbar) and \citet{noll86} (3 mbar). 
We find about 1 order of magnitude
less \acet\ at 10~mbar than \citet{clarke82} and \citet{gladstone83},
though 1 order of magnitude greater than \citet{bezard95}. Note that
the prior results at 10~mbar are mutually inconsistent. The reason for
this is unclear, but the lower abundance compared to \ethane\ may be
partly responsible. At 1 mbar,
our result appears to be compatible with \citet{bezard95}, if some
error is allowed on those results.

\subsection{Chemistry and Dynamics}
\label{sect:photochem}

The meridional distribution of a gas species depends upon the relative
chemical and transport lifetimes. In the absence of meridional
transport, for a species that is short-lived photochemically (compared to
seasonal time-scales) we expect that the observed distributions in
latitude should, to first order, follow the incident seasonal
insolation. This was noted in Paper I for \acet , whose lifetime is
short ($3\times 10^{7}$~s at 5~mbar) compared to both a Jovian season
($1\times 10^{8}$~s), and the inferred timescale for meridional mixing
from analysis of Comet Shoemaker-Levy 9 debris \citep[$6\times
  10^{8}$~s,][]{lellouch02}.

For species with photochemical lifetimes approaching
seasonal time-scales, a phase lag in response to the insolation is
expected with a diminished peak to peak variation in the mixing
ratio. And finally for long lived species (e.g. ethane, with lifetime
$3\times 10^{10}$~s at 5~mbar) only small perturbations
about annual global averages are expected. Meridional transport will
wash out any differences. Additionally the vertical transport
required to close the meridional cells will cause changes in the
mixing ratios at various pressures depending upon the vertical profile
of the mixing ratio, e.g. vertical transport bringing down \acet\ rich
air to a given pressure level. 

Figure \ref{fig:lifetimes} compares the photochemical lifetimes (halving and
doubling) for \acet\ and \ethane\ from the 1-D model calculations described
in Paper I to the Jovian year, the solar cycle, and the eddy
mixing time (H$^2$/K, where H is the mixed atmospheric scale height and K
is the eddy diffusion coefficient). The acetylene lifetimes are less
than the seasonal, solar cycle, and eddy mixing times, while the
opposite is the case for ethane. In the model both the acetylene and
ethane mixing ratios increase with decreasing pressure, so downward
transport of air parcels will increase their relative abundances and
conversely for rising parcels. However, ethane is much more uniformly
mixed with height due to its chemical lifetimes being longer than the
eddy transport time. Thus vertical transport will have more of an
effect on the \acet\ mixing ratio than for ethane, as enriched air is
downwardly advected.

\begin{figure}[h]
\caption{Appears Here.}
\label{fig:lifetimes}
\end{figure}

For acetylene and equinoctial conditions, we expect a more symmetric
distribution compared to solsticial. Our results seem to validate this
hypothesis, as the Voyager distribution of \acet\ appears much more
symmetric than the Cassini distribution, which is biased towards the
northern hemisphere - as expected for northern fall equinox and summer
solstice conditions respectively. For ethane, the reverse is seen: the
Cassini (solstice) epoch shows a more uniform latitude distribution
than the Voyager (equinox) epoch, which may indicate a phase lag
between dynamics and seasons. 

The higher abundances of \ethane\ seen at
most latitudes by CIRS may in part be due to the increased mean
insolation at northern solstice relative to fall equinox: Jupiter was
5\% closer to the Sun in December 2000 compared to March 1979,
resulting in 10\% higher solar flux (21\% flux variation possible
due to solar distance over an entire year). 
In addition, we note that the increase in \ethane\ towards the poles
at both epochs may be partially due to particle-induced auroral
chemistry \citep{yelle01}.

\subsection{Comparison to Saturn}
\label{sect:saturn}

Our findings for Jupiter closely parallel recent studies of the minor
hydrocarbons in Saturn's atmosphere, which have mostly found a
decrease in acetylene and constant or increasing ethane from equator
to pole. \citet{howett07}, using Cassini CIRS data,
found a decrease in the abundance of \acet\
at 2~mbar from 30--68\dg S by a factor 1.8, although there was an
initial increase from 18--30\dg S. Ethane on the other hand increased
from the equator to the south pole by 2.5. \citet{hesman09}, combining
results from Cassini CIRS and ground based spectroscopy with the IRTF
and Celeste, found a decrease in \acet\ by a factor 2.7 from 5--75\dg
S, although there was a rise again at 87\dg S. Ethane was constant
from equatorial to southern mid-latitudes, where it began to rise in
abundance, finally doubling at the south pole. 

Both these studies used nadir-geometry spectra, limiting the range of
vertical sensitivity. \citet{guerlet09} however have analyzed Cassini
CIRS limb spectra to yield a more detailed vertical picture of the
hydrocarbon meridional abundance variations. Acetylene is found to be
sharply peaked at the equator at 1 mbar, with a gradual decrease
through southern and northern mid-latitudes, while ethane is nearly
flat from 50\dg N--50\dg S, where is rises towards the south pole.

Overall, we see some intriguing similarities between the meridional
variations of the secondary hydrocarbons on Jupiter and
Saturn, although the details differ. The implication is that
researches into understanding the origins of these distributions must
proceed in parallel, and advances in understanding the mechanisms of
one body are likely to improve our appreciation of the other. Note
that on Saturn however we must be mindful of the effect of the ring
shadow,  which is seasonally varying over a 29.5 (earth) year
period. This has the effect of shielding the low latitudes, reducing
both the photolysis of methane, and the higher hydrocarbons, which
must be accounted for correctly in photochmeical models.

%%%%%%%%%%%%%%%%%%%%%%%%%%%%%%%%%%%%%%%%%%%%%%%%%%%%%%%%%%%%%%%%%%%%%%%%%
%%%%%%%%%%%%%%%%%%%%%%%%%%%%%%%%%%%%%%%%%%%%%%%%%%%%%%%%%%%%%%%%%%%%%%%%%
\section{Summary and Conclusions}
\label{conclusions}

In this paper we have analyzed two infrared spectral datasets of
Jupiter: (i) the Voyager IRIS dataset, pertaining to the northern fall
equinox of 1979, and (ii) the Cassini CIRS dataset, taken near the
northern summer solstice of 2000. We have modeled the spectra to retrieve,
firstly atmospheric temperatures through the $\nu_4$ band of \methane\
at $\sim1305$\cm , and secondly the abundances of \ethane\ and \acet\
based on the derived temperatures. Our results extend from the upper
troposphere to the upper stratosphere for temperature and \acet\
abundance, and cut-off slightly lower in the middle stratosphere for
\ethane . The results comprise the first comprehensive 2-D comparison 
of the abundances of \acet\ and \ethane\ in Jupiter's atmosphere at
two different seasons. The results have been archived on-line for
reference, as described in \S \ref{sect:results}.

Our new findings confirm, improve and extend the results of our
previous paper \citep{nixon07}. The distribution of \acet\ at the
Cassini epoch is confirmed, whilst that of \ethane\ is corrected,
largely due to revised spectral line intensities published in the
interim \citep{auwera07}. We have also extended the Cassini
distributions to a second Jovian season (Voyager), showing important
similarities and differences. 

For ethane, the 2-D distribution is
quite similar, although we see that the Cassini picture is more
symmetric than that of Voyager, where \ethane\ is somewhat depleted in
the north relative to the south. Also, the Cassini abundances are
mostly higher overall, perhaps due to smaller solar distance.
For acetylene the difference is much more dramatic, showing a uniform
meridional profile at equinox, but strong depletions towards both
poles at solstice. Therefore the solstice distribution seems to match
the annual-mean picture that would be predicted by a 1-D photochemical
model.

It is clear that Jupiter's stratosphere is indeed a complex
environment, where chemistry is coupled to dynamics in ways that are
poorly understood at present. Much additional modeling work is
required before these seasonal changes can be confidently interpreted,
and comprehensive radiative-dynamical-chemical 
models in two or three dimensions
are urgently needed. The rewards of successful modeling of this data
are large, and include the significant advancement of our
understanding not only of the Jovian atmosphere, but also that of Saturn
and the other outer planets as well.

%%%%%%%%%%%%%%%%%%%%%%%%%%%%%%%%%%%%%%%%%%%%%%%%%%%%%%%%%%%%%%%%%%%%%%%%%%%
%%%%%%%%%%%%%%%%%%%%%%%%%%%%%%%%%%%%%%%%%%%%%%%%%%%%%%%%%%%%%%%%%%%%%%%%%%%

\clearpage
{\bf Acknowledgements}

The acquisition of CIRS data is the result of the collective efforts of a
large number of people, including the following who worked on various
aspects of CIRS science planning, instrument commanding, uplink,
calibration and databasing: S.B. Calcutt, R.C. Carlson, M.H. Elliott,
E. Guandique, M. Kaelberer, V.G. Kunde, E. Lellouch, A. Mamoutkine,
P.J. Schinder, M.E. Segura, J.S. Tingley, and also many engineers and science
planners at the Jet Propulsion Laboratory. We would like to thank
R.A. West, A.J. Friedson and Y.L. Yung for helpful discussions during
the preparation of this manuscript. During the research for and
writing of this report, RKA and FMF were funded by the NASA Cassini
Project, and CAN and MA were supported by the NASA Outer Planets Research
Program. Portions of this work were carried out by the Jet Propulsion
Laboratory, under contract with the National Aeronautics and Space
Administration. 

\appendix

%\section{Appendix A}
%\label{app:one}

\clearpage

\bibliographystyle{elsarticle-harv}
%\bibliographystyle{harvard}
%\bibliography{/home/nixon/tex/CIRS/cirs_jup}

%%%%%%%%%%%%%%%%%%%%%%%%%%%%%%%%% TABLES %%%%%%%%%%%%%%%%%%%%%%%%%%%%%%

\def\baselinestretch{0.8}

\clearpage
\thispagestyle{empty}

\begin{table}[p]
\begin{centering}
\footnotesize
\begin{tabular}{cccccc}
\multicolumn{6}{c}{\bf TABLE I} \\
\multicolumn{6}{c}{\bf Cassini CIRS Observational Data} \\
\hline
Mean     & No. of FP3 & No. of FP4 & Mean Emis. & NESR & NESR \\
Latitude & Spectra  & Spectra  & Angle (\dg) &  760 \cm\ & 1300 \cm\ \\
\hline
-66.30 &   19 &    7 &  71.21 & $3.50 \times 10^{ -9}$ & $8.50 \times 10^{-10}$ \\
-62.24 &   73 &   67 &  67.76 & $1.92 \times 10^{ -9}$ & $3.30 \times 10^{-10}$ \\
-57.28 &  172 &  139 &  64.13 & $1.50 \times 10^{ -9}$ & $2.30 \times 10^{-10}$ \\
-52.49 &  231 &  234 &  60.81 & $1.36 \times 10^{ -9}$ & $1.89 \times 10^{-10}$ \\
-47.54 &  306 &  313 &  56.93 & $1.09 \times 10^{ -9}$ & $1.71 \times 10^{-10}$ \\
-42.31 &  559 &  332 &  52.92 & $0.94 \times 10^{ -9}$ & $1.71 \times 10^{-10}$ \\
-37.36 &  855 &  366 &  48.96 & $0.94 \times 10^{ -9}$ & $1.74 \times 10^{-10}$ \\
-32.48 & 1171 &  402 &  45.76 & $1.14 \times 10^{ -9}$ & $1.71 \times 10^{-10}$ \\
-27.84 &  644 &  377 &  38.73 & $1.07 \times 10^{ -9}$ & $1.71 \times 10^{-10}$ \\
-22.32 &  446 &  523 &  36.29 & $1.07 \times 10^{ -9}$ & $1.93 \times 10^{-10}$ \\
-17.37 &  427 &  501 &  32.23 & $1.02 \times 10^{ -9}$ & $1.65 \times 10^{-10}$ \\
-12.20 &  471 & 2144 &  34.77 & $1.01 \times 10^{ -9}$ & $1.16 \times 10^{-10}$ \\
 -7.93 &  437 & 1149 &  27.51 & $1.07 \times 10^{ -9}$ & $1.59 \times 10^{-10}$ \\
 -2.65 &  442 &  503 &  23.41 & $1.07 \times 10^{ -9}$ & $2.07 \times 10^{-10}$ \\
  2.50 &  505 &  551 &  21.89 & $1.01 \times 10^{ -9}$ & $1.83 \times 10^{-10}$ \\
  7.55 &  413 &  536 &  22.49 & $1.08 \times 10^{ -9}$ & $1.88 \times 10^{-10}$ \\
 12.56 &  580 &  558 &  26.31 & $1.07 \times 10^{ -9}$ & $2.02 \times 10^{-10}$ \\
 17.52 &  604 &  568 &  28.16 & $1.03 \times 10^{ -9}$ & $2.05 \times 10^{-10}$ \\
 22.44 &  525 &  492 &  29.66 & $1.09 \times 10^{ -9}$ & $2.20 \times 10^{-10}$ \\
 27.33 &  478 &  568 &  34.89 & $1.08 \times 10^{ -9}$ & $2.52 \times 10^{-10}$ \\
 32.47 &  464 &  548 &  38.67 & $1.09 \times 10^{ -9}$ & $2.39 \times 10^{-10}$ \\
 37.51 &  491 &  558 &  41.09 & $1.17 \times 10^{ -9}$ & $2.11 \times 10^{-10}$ \\
 42.56 &  430 &  489 &  46.05 & $1.14 \times 10^{ -9}$ & $2.13 \times 10^{-10}$ \\
 47.37 &  437 &  400 &  50.57 & $1.16 \times 10^{ -9}$ & $2.22 \times 10^{-10}$ \\
 52.29 &  381 &  402 &  55.60 & $1.15 \times 10^{ -9}$ & $2.30 \times 10^{-10}$ \\
 57.16 &  334 &  294 &  59.12 & $1.23 \times 10^{ -9}$ & $2.51 \times 10^{-10}$ \\
 62.24 &  220 &  187 &  63.54 & $1.43 \times 10^{ -9}$ & $3.10 \times 10^{-10}$ \\
 67.14 &   96 &   97 &  66.42 & $1.94 \times 10^{ -9}$ & $3.79 \times 10^{-10}$ \\
\hline
\end{tabular}
\normalsize
\newline
N. B. this table, included here for completeness, is identical to
TABLE I of \citet{nixon07}.
\\
\end{centering}
\label{tab:highresspecs}
\end{table}

\clearpage
\thispagestyle{empty}

\begin{table}[p]
\begin{centering}
\footnotesize
\begin{tabular}{cccc}
\multicolumn{4}{c}{\bf TABLE II} \\
\multicolumn{4}{c}{\bf Voyager IRIS Observational Data} \\
\hline
Mean Latitude & No. of spectra & Mean Emission Angle & NESR$^{\dagger}$ \\
-59.97 & 27 & 60.39 & $9.62\times10^{-10}$ \\
-55.02 & 32 & 54.65 & $8.84\times10^{-10}$ \\
-49.49 & 43 & 49.76 & $7.62\times10^{-10}$ \\
-45.04 & 55 & 45.34 & $6.74\times10^{-10}$ \\
-40.73 & 46 & 40.43 & $7.37\times10^{-10}$ \\
-34.29 & 52 & 34.06 & $6.93\times10^{-10}$ \\
-30.34 & 62 & 30.49 & $6.35\times10^{-10}$ \\
-25.01 & 53 & 25.33 & $6.87\times10^{-10}$ \\
-19.49 & 63 & 20.91 & $6.30\times10^{-10}$ \\
-15.00 & 68 & 19.52 & $6.06\times10^{-10}$ \\
-9.59 & 74 & 17.16 & $5.81\times10^{-10}$ \\
-4.80 & 88 & 15.22 & $5.33\times10^{-10}$ \\
-0.34 & 79 & 12.62 & $5.63\times10^{-10}$ \\
4.79 & 73 & 10.32 & $5.85\times10^{-10}$ \\
9.72 & 76 & 10.56 & $5.74\times10^{-10}$ \\
14.82 & 65 & 12.01 & $6.20\times10^{-10}$ \\
20.02 & 55 & 16.67 & $6.74\times10^{-10}$ \\
25.19 & 58 & 21.55 & $6.57\times10^{-10}$ \\
30.01 & 60 & 26.04 & $6.45\times10^{-10}$ \\
35.05 & 58 & 30.50 & $6.57\times10^{-10}$ \\
39.91 & 59 & 35.87 & $6.51\times10^{-10}$ \\
44.04 & 48 & 40.10 & $7.22\times10^{-10}$ \\
50.05 & 40 & 45.35 & $7.91\times10^{-10}$ \\
54.35 & 37 & 49.86 & $8.22\times10^{-10}$ \\
58.63 & 22 & 54.34 & $1.07\times10^{-09}$ \\
\hline
\end{tabular}
\normalsize
\newline
$^{\dagger}$ NESR = Noise Equivalent Spectral Radiance (1-$\sigma$) in
units of \radunit . \\
\end{centering}
\label{tab:obs}
\end{table}

\clearpage
\thispagestyle{empty}

\begin{table}[p]
\begin{centering}
\footnotesize
\begin{tabular}{lllccc}
\multicolumn{6}{c}{\bf TABLE III} \\
\multicolumn{6}{c}{\bf Comparison of Altitude Sensitivity: Equator} \\
\hline
 & &  & & Lower & Upper \\
Wavenumber & Primary & Retrieved & Troposphere & Stratosphere & Stratosphere \\
{\cm} & Opacity & Parameter & Range (mbar) & Range (mbar) & Range (mbar) \\
\hline
 625 &  {\hydrogen} & Temp & {\em  98--339 (206)} &  &  \\ 
 & & & {\bf 106--339 (206)} &  &  \\ 
 722 &      {\acet} &  VMR & {\em  74--339 (178)} & {\em  0.93--31.27 ( 9.80)} &  \\ 
 & & & {\bf 266--748 (516)} & {\bf  0.12-- 7.21 ( 3.79)} &  \\ 
 729 &      {\acet} &  VMR & {\em  84--374 (219)} & {\em  0.03--28.39 ( 4.59)} &  \\ 
 & & & {\bf 262--772 (525)} & {\bf  0.17-- 7.45 ( 3.79)} &  \\ 
 731 &      {\acet} &  VMR & {\em  87--392 (242)} & {\em  0.01-- 2.83 ( 0.15)} &  \\ 
 & & & {\bf 262--772 (533)} & {\bf  0.17-- 7.45 ( 3.85)} &  \\ 
 822 &    {\ethane} &  VMR & {\em 137--597 (356)} & {\em  0.70--13.97 ( 4.38)} &  \\ 
 & & & {\bf  73--679 (432)} & {\bf  0.86--10.45 ( 4.38)} &  \\ 
1250 &   {\methane} & Temp &  & {\em  1.59--21.59 ( 5.95)} &  \\ 
 & & &  & {\bf  2.37--15.39 ( 6.14)} &  \\ 
1304 &   {\methane} & Temp &  & {\em  0.92-- 7.21 ( 3.27)} &  \\ 
 & & &  & {\bf  1.01-- 7.69 ( 3.61)} &  \\ 
1306 &   {\methane} & Temp &  & {\em  0.40-- 2.79 ( 1.25)} & {\em 0.003--0.025 (0.006)} \\ 
 & & &  & {\bf  0.76-- 9.96 ( 3.12)} & {\bf 0.003--0.019 (0.007)} \\ 
\hline
\end{tabular}
\normalsize
\newline
VMR = Volume Mixing Ratio. \\
{\em Italic text} is Cassini CIRS; {\bf bold text} is Voyager IRIS. \\
In each case the range and peak (in parentheses) of the sensitivity is given.\\
\end{centering}
\label{tab:fds_equat}
\end{table}

\clearpage
\thispagestyle{empty}

\begin{table}[p]
\begin{centering}
\footnotesize
\begin{tabular}{lllccc}
\multicolumn{6}{c}{\bf TABLE IV} \\
\multicolumn{6}{c}{\bf Comparison of Altitude Sensitivity: North} \\
\hline
 & &  & & Lower & Upper \\
Wavenumber & Primary & Retrieved & Troposphere & Stratosphere & Stratosphere \\
{\cm} & Opacity & Parameter & Range (mbar) & Range (mbar) & Range (mbar) \\
\hline
 625 &  {\hydrogen} & Temp & {\em  63--254 (159)} &  &  \\ 
 & & & {\bf  66--262 (156)} &  &  \\ 
 722 &      {\acet} &  VMR & {\em  71--284 (106)} & {\em  3.22--39.19 (16.95)} &  \\ 
 & & & {\bf 206--587 (399)} & {\bf  1.51--11.15 ( 5.31)} &  \\ 
 729 &      {\acet} &  VMR & {\em  74--323 (111)} & {\em  0.09--39.19 (17.23)} &  \\ 
 & & & {\bf 202--597 (399)} & {\bf  1.54--11.89 ( 5.48)} &  \\ 
 731 &      {\acet} &  VMR & {\em  73--392 (242)} & {\em  0.06-- 2.83 ( 0.16)} &  \\ 
 & & & {\bf 202--597 (399)} & {\bf  1.54--12.08 ( 5.57)} &  \\ 
 822 &    {\ethane} &  VMR & {\em  90--525 (318)} & {\em  1.09--19.91 ( 2.83)} &  \\ 
 & & & {\bf 135--551 (345)} & {\bf  2.70--21.59 ( 9.04)} &  \\ 
1250 &   {\methane} & Temp &  & {\em  2.09--25.77 ( 9.64)} &  \\ 
 & & &  & {\bf  2.37--21.59 ( 7.95)} &  \\ 
1304 &   {\methane} & Temp &  & {\em  1.19-- 6.44 ( 2.97)} &  \\ 
 & & &  & {\bf  0.77-- 6.99 ( 2.88)} &  \\ 
1306 &   {\methane} & Temp &  & {\em  0.39-- 2.45 ( 1.31)} & {\em 0.002--0.007 (0.004)} \\ 
 & & &  & {\bf  0.59--14.43 ( 2.30)} & {\bf 0.003--0.015 (0.005)} \\ 
\hline
\end{tabular}
\normalsize
\newline
VMR = Volume Mixing Ratio. \\
{\em Italic text} is Cassini CIRS; {\bf bold text} is Voyager IRIS. \\
In each case the range and peak (in parentheses) of the sensitivity is given.\\
\end{centering}
\label{tab:fds_north}
\end{table}

\clearpage
\thispagestyle{empty}

\begin{table}[p]
\begin{centering}
\footnotesize
\begin{tabular}{rrrrrrrr}
\multicolumn{8}{c}{\bf TABLE V} \\
\multicolumn{8}{c}{\bf Comparison to Previous Results} \\
\hline
Gas & Ref. & Press. & Lat.$^{\dagger}$ & Date & Mole & Voyager 79 & Cassini
2000 \\
 Species   &      & (mbar) &    & & Frac. & Mole Frac. & Mole Frac. \\
\hline
\ethane\ & F81 & 0.005 & 16.5\dg N & 07/79 & $(2.5\pm 1.7)$ ppm & N/A & --- \\
 & G83 & 4 & 30\dg S-30\dg N & 12/78 & $(6.6\pm 5.3)$ ppm & 
$(2.3\pm0.7)$ ppm & --- \\
 & & & & \& 6/79 & & & \\
  & N86 & 5 & 8\dg S-8\dg N & 06/80 & $(2.6\pm 0.9)$ ppm 
& $(1.9\pm 0.6)$ ppm & --- \\
 & L93 & 0.3--3 & 45--55\dg N & 12/89 & $(3.6\pm 0.9)$ ppm & --- &
$(9.4\pm 1.8)$ ppm$^{\ddagger}$ \\
 & & & & & & & \\
 \acet\ & O80 & 10 & 30\dg S-30\dg N & 1/79 & 19 ppb &
 $(13\pm 5)$ ppb & --- \\
 & & & & \& 4/79 & & & \\
& C82 & 10 & 30\dg S-30\dg N & 1978--1980 & $(100\pm 50)$ ppb &
 $(13\pm 5)$ ppb & --- \\
& G83 & 10 & 30\dg S-30\dg N & 12/78  & $(100\pm 10)$ ppb &
 $(13\pm 5)$ ppb & --- \\
 & & & & \& 6/79 & & & \\
& W85 & 30 & 30\dg S-30\dg N & 1978-1980 & $(30\pm 10)$ ppb &
 N/A & --- \\
& N86 & 3 & 8\dg S-8\dg N & 6/80 & $(34\pm 17)$ ppb & 
 $(62\pm 20)$ ppb & --- \\
& B95 & 0.1 & 10--36\dg S & 1/91 & 4 ppm  & N/A & --- \\
& B95 & 1  & 10--36\dg S & 1/91 & 100 ppb  & 
 $(250 \pm 100)$ ppb & --- \\
& B95 & 10 & 10--36\dg S & 1/91 & 1 ppb  & 
 $(16\pm 6)$ ppb & --- \\
\hline
\end{tabular}
\normalsize
\newline
$^{\dagger}$Voyager/Cassini data calculated at mean latitude of range,
where applicable. $^{\ddagger}$At 3 mbar. References: O80 =
\citet{owen80}; F81 = \citet{festou81}; C82 = \citet{clarke82}; G83 =
\citet{gladstone83}; W85 = \citet{wagener85}; N86 = \citet{noll86};
L93 = \citet{livengood93}; B95 = \citet{bezard95}. \\
\end{centering}
\label{tab:comp}
\end{table}

%%%%%%%%%%%%%%%%%%%%%%%%%%%%% FIGURE CAPTIONS %%%%%%%%%%%%%%%%%%%%%%%%%%%%
\def\baselinestretch{1.6}

\clearpage

{\bf Figure Captions}

Fig. \ref{fig:gas_apriori} 
Initial assumed {\em a priori} temperature (left) and gas profiles (right)
for Jupiter atmospheric model.

Fig. \ref{fig:temp_fds} 
Functional derivatives for temperature from the Voyager IRIS spectral
data: (a) from the \hydrogen\
continuum region giving tropospheric information, and (b) from the
\methane\ $\nu_4$ band, sensitive to the lower stratosphere near 5 mbar
and the upper stratosphere at 5 $\mu$bar.

Fig. \ref{fig:c2h2_fds} 
Functional derivative for gas abundance from the \acet\ $\nu_5$
band. Note the emission emanating from the stratosphere at 1 mbar,
and also tropospheric absorption at $\sim$200~mbar. Some
information is available from the upper stratosphere at
10--100~\microbar\ due to the Q-branch (730 \cm ). 

Fig. \ref{fig:c2h6_fds} 
Functional derivative for gas abundance from the \ethane\ $\nu_9$
band. Note the emission emanating from the stratosphere at 5 mbar,
and also tropospheric absorption at $\sim$200~mbar. 

Fig. \ref{fig:comp_fds_equat} 
Examples of the FWHM sensitivity (symbols at bar ends) and peak
sensitivity for the contribution functions at various representative
wavenumbers, comparing Cassini ($\times$) to Voyager
($\diamond$), computed for Jupiter's equator. 
Even though the spectral resolutions are quite different
(0.5 \cm versus 4.3 \cm ), the vertical information regions are
similar except for the \acet\ hotband near 731 \cm\ for which the peak
has moved considerably lower for IRIS due to lower spectral resolution.

Fig. \ref{fig:comp_fds_north} 
As in previous figure, but computed for 58\dg N.

Fig. \ref{fig:nlte} 
Comparison of the vertically-varying collisional relaxation time
($\tau_{\rm C}$) for the $\nu_4$ band of \methane\ to the spontaneous
emission timescale $\tau_{\rm A}$, assumed 0.4~s following
\citet{halthore94}. The break-down of LTE conditions is computed to be at
$\sim 0.8$~\microbar .

Fig. \ref{fig:temp_contours} 
Retrieved temperatures from Voyager IRIS data (a) and Cassini CIRS
(b). (c) shows the difference: Cassini-Voyager. This may be compared
to Simon-Miller et al. (2006) Fig. 2. We find that the stratosphere at
$\sim$2 mbar is warmer in the north in 2000 compared to 1979, and at
0.01 mbar is cooler.

Fig. \ref{fig:c2h2_contours} 
Comparison of retrieved abundances for \acet\ at the (a) Voyager and
(b) Cassini epochs. The vertical range has been restricted to match
the smaller span (IRIS) (see Figs. 
\ref{fig:comp_fds_equat} and \ref{fig:comp_fds_north}). Abundances in
the lower stratosphere ($\sim$ 10 mbar) have dropped substantially in
2000 at the poles relative to 1979.

Fig. \ref{fig:c2h6_contours} 
Comparison of retrieved abundances for \ethane\ at the (a) Voyager and
(b) Cassini epochs. The changes seen are much less dramatic than for
\acet , nevertheless the Cassini (2000) stratospheric abundances
appear to be higher and more symmetric in latitude compared to the
Voyager (1979) contours, which show a positive bias to southern hemisphere.

Fig. \ref{fig:cirs_comp}
Comparison of the retrieved abundances of \acet\ and \ethane\ from the
Cassini CIRS dataset, taking the difference between the 2007 and 2010
retrievals. The differences for acetylene, due mainly to changes in
the instrumental lineshape assumed (see text for discussion), are
typically 1--10\% across most of the range (c.f. formal retrieval
errors of $\sim 40--50$\%), reaching maximum divergences of +20\% and
-40\% in the northern stratosphere and troposphere respectively, which
approach the size of the inferred random errors. For ethane, the
differences are greater, with large changes (-40 to -60\%) occuring in
the mid-stratosphere at 10~mbar at the peak of the contribution
function, and are significantly larger than the random errors
(15--20\%). This is due to the revised $\nu_9$ band intensities 
used for \ethane , as discussed in \S 3.2. 

Fig. \ref{fig:lifetimes} Photochemical production (solid color lines) and
loss (broken color lines) times (time for mixing ratio of species to
increased/decreased by a factor of two) for \acet\ (red) and \ethane\
(blue) for the model used in \citet{nixon07}. These are compared to
the eddy mixing time, the Jovian year, and half the period of the solar cycle.

%%%%%%%%%%%%%%%%%%%%%%%%%%%%%%%%%%%%%%%%%%%%%%%%%%%%%%%%%%%%%%%%%%%%%
%%%%%%%%%%%%%%%%%%%%%%%%% ACTUAL FIGURES %%%%%%%%%%%%%%%%%%%%%%%%%%%%
%%%%%%%%%%%%%%%%%%%%%%%%%%%%%%%%%%%%%%%%%%%%%%%%%%%%%%%%%%%%%%%%%%%%%

\LARGE

\clearpage

\begin{figure}[t]\centerline{
\epsfig{file=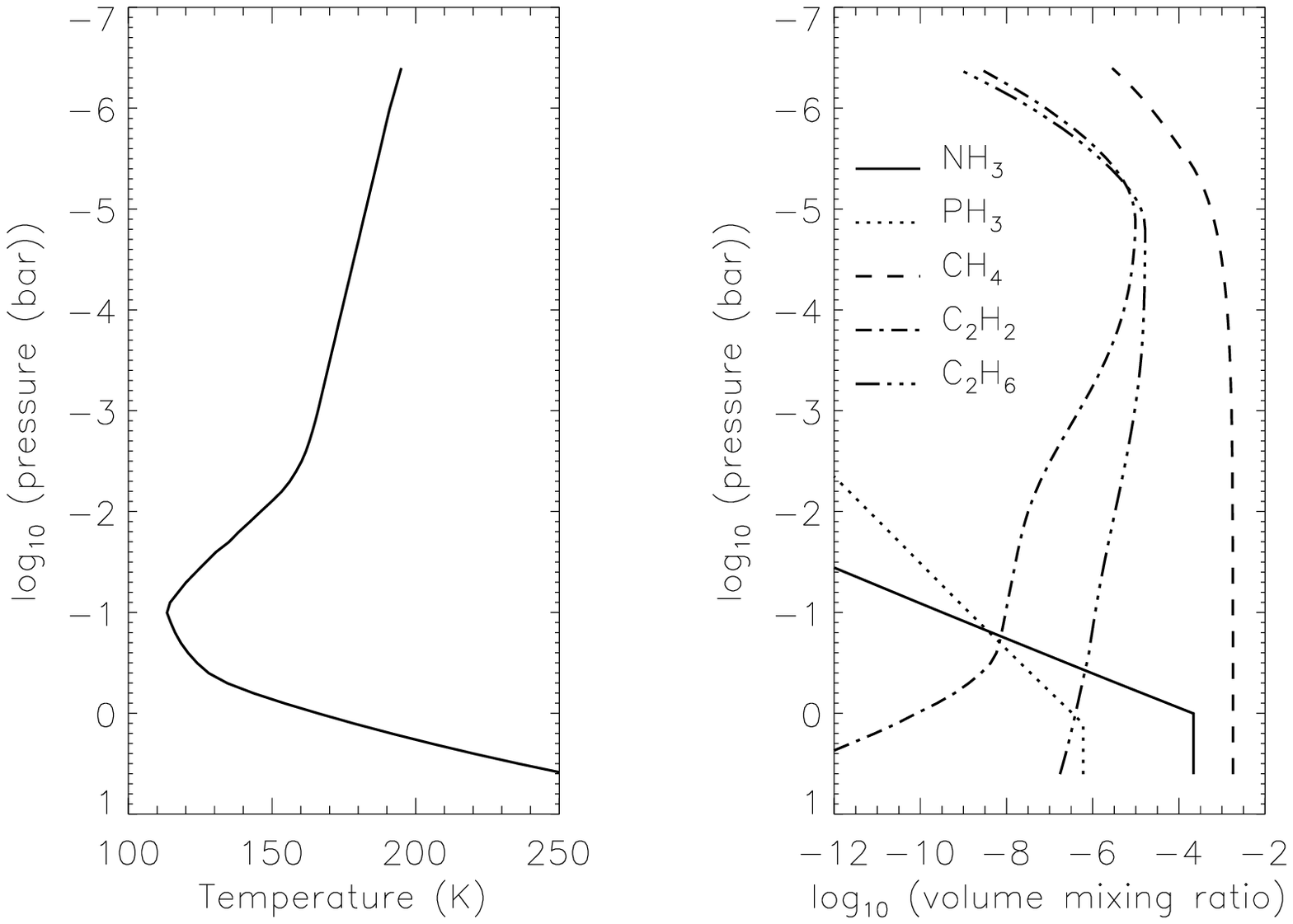, height=13cm, angle=0}
}
\end{figure}

Nixon et al. {\bf Figure \ref{fig:gas_apriori}}

\clearpage

\begin{figure}[t]\centerline{
\epsfig{file=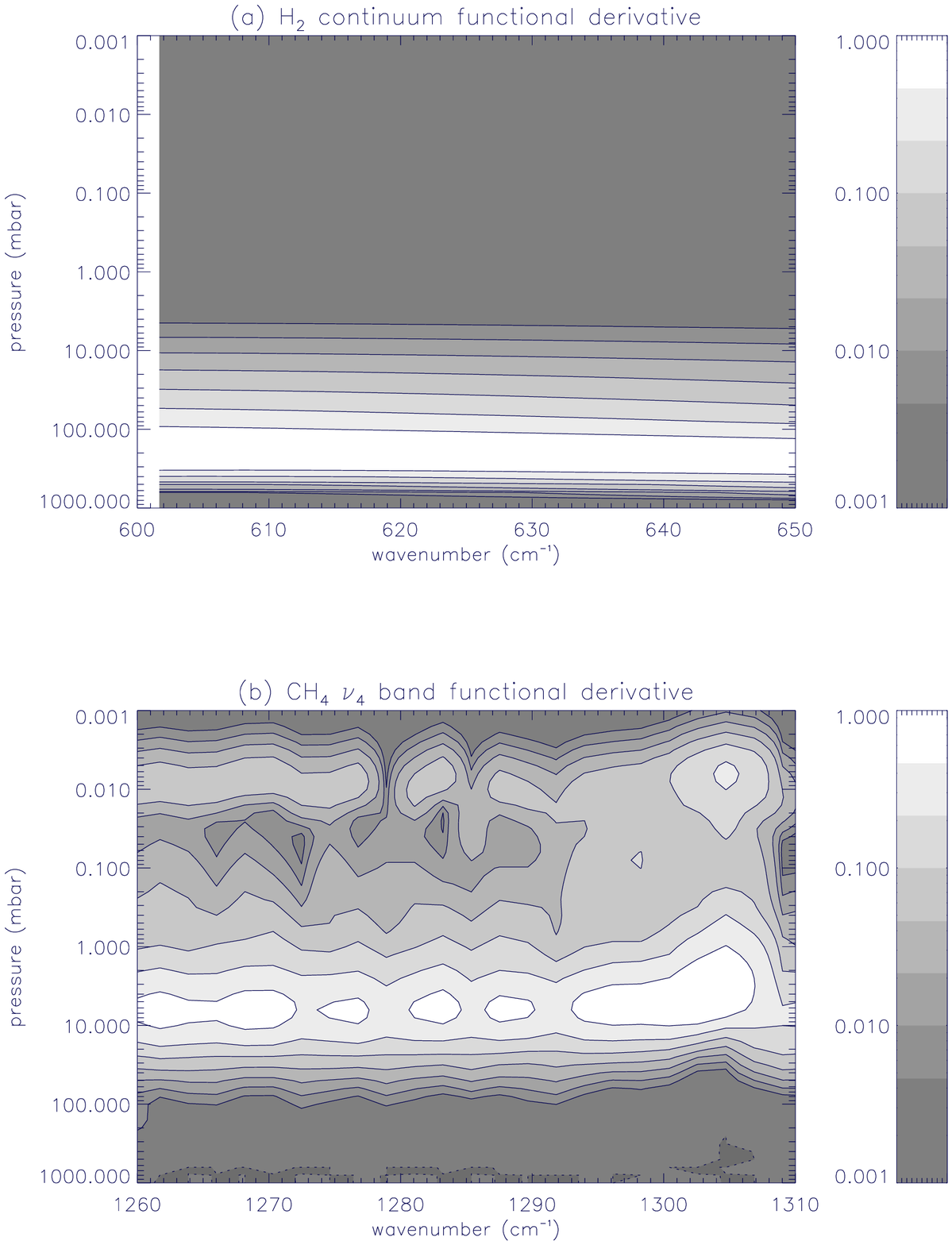, height=19cm, angle=0}
}
\end{figure}

Nixon et al. {\bf Figure \ref{fig:temp_fds}}

\clearpage

\begin{figure}[t]\centerline{
\epsfig{file=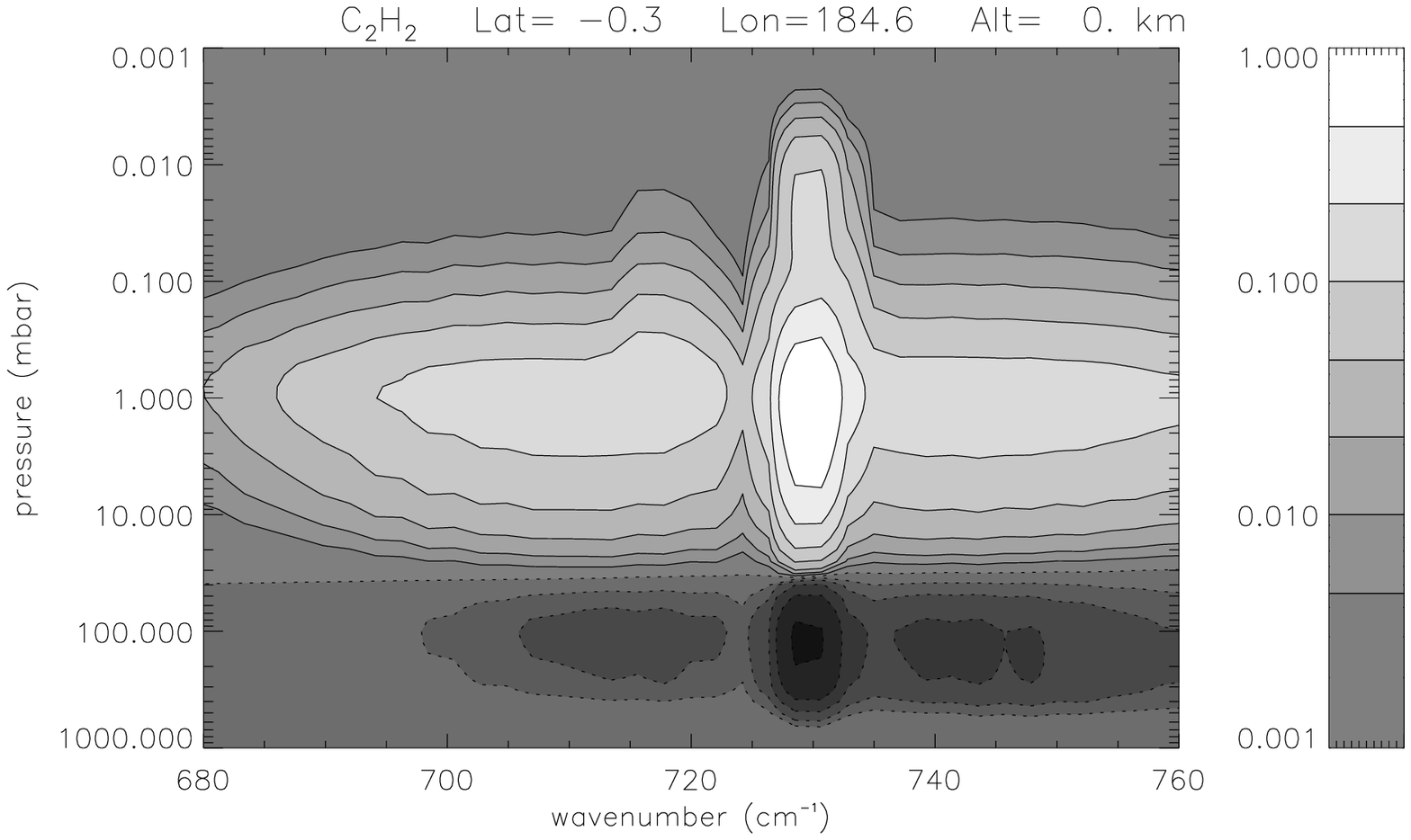, height=19cm, angle=0}
}
\end{figure}

Nixon et al. {\bf Figure \ref{fig:c2h2_fds}}

\clearpage

\begin{figure}[t]\centerline{
\epsfig{file=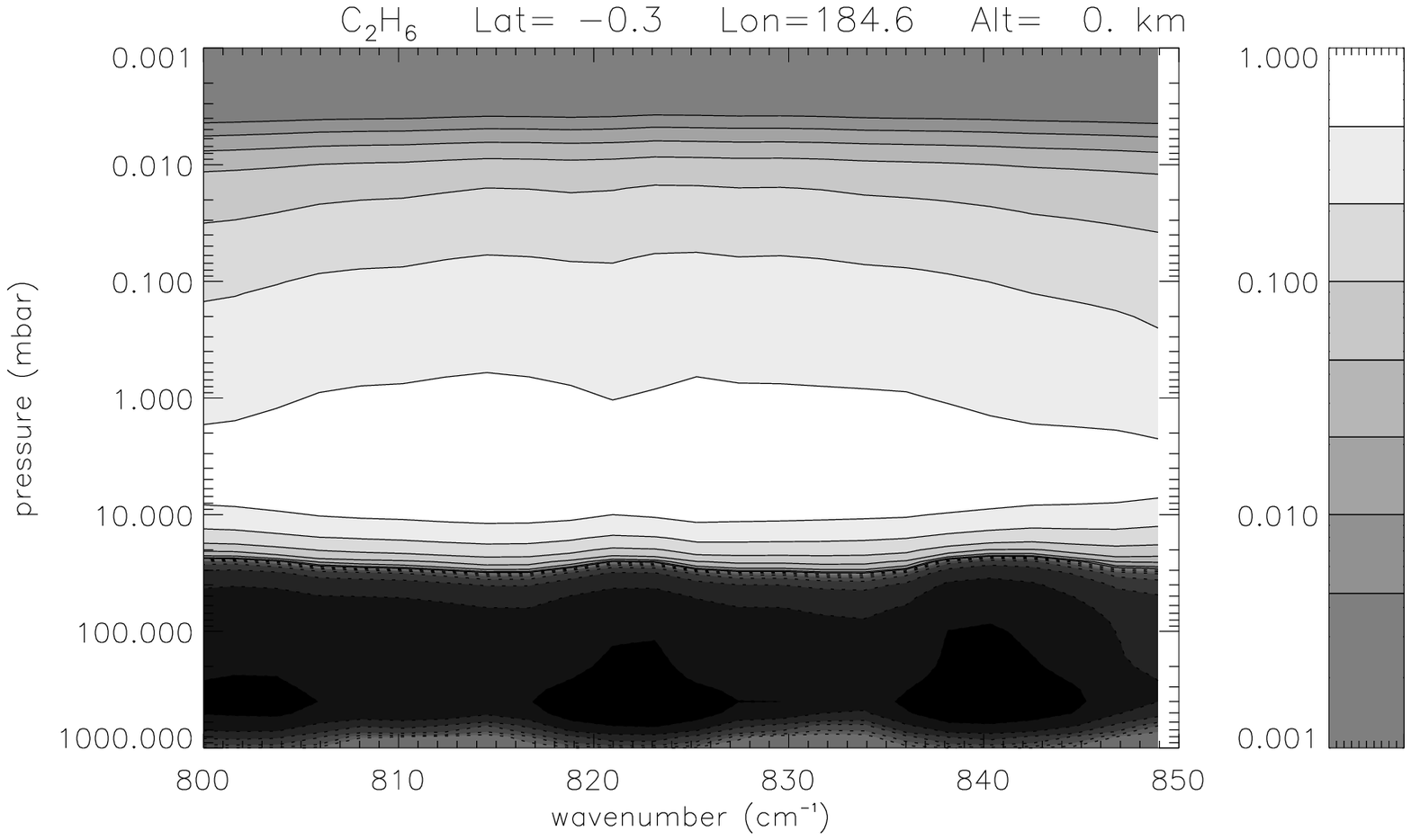, height=19cm, angle=0}
}
\end{figure}

Nixon et al. {\bf Figure \ref{fig:c2h6_fds}}

\clearpage

\begin{figure}[t]\centerline{
\epsfig{file=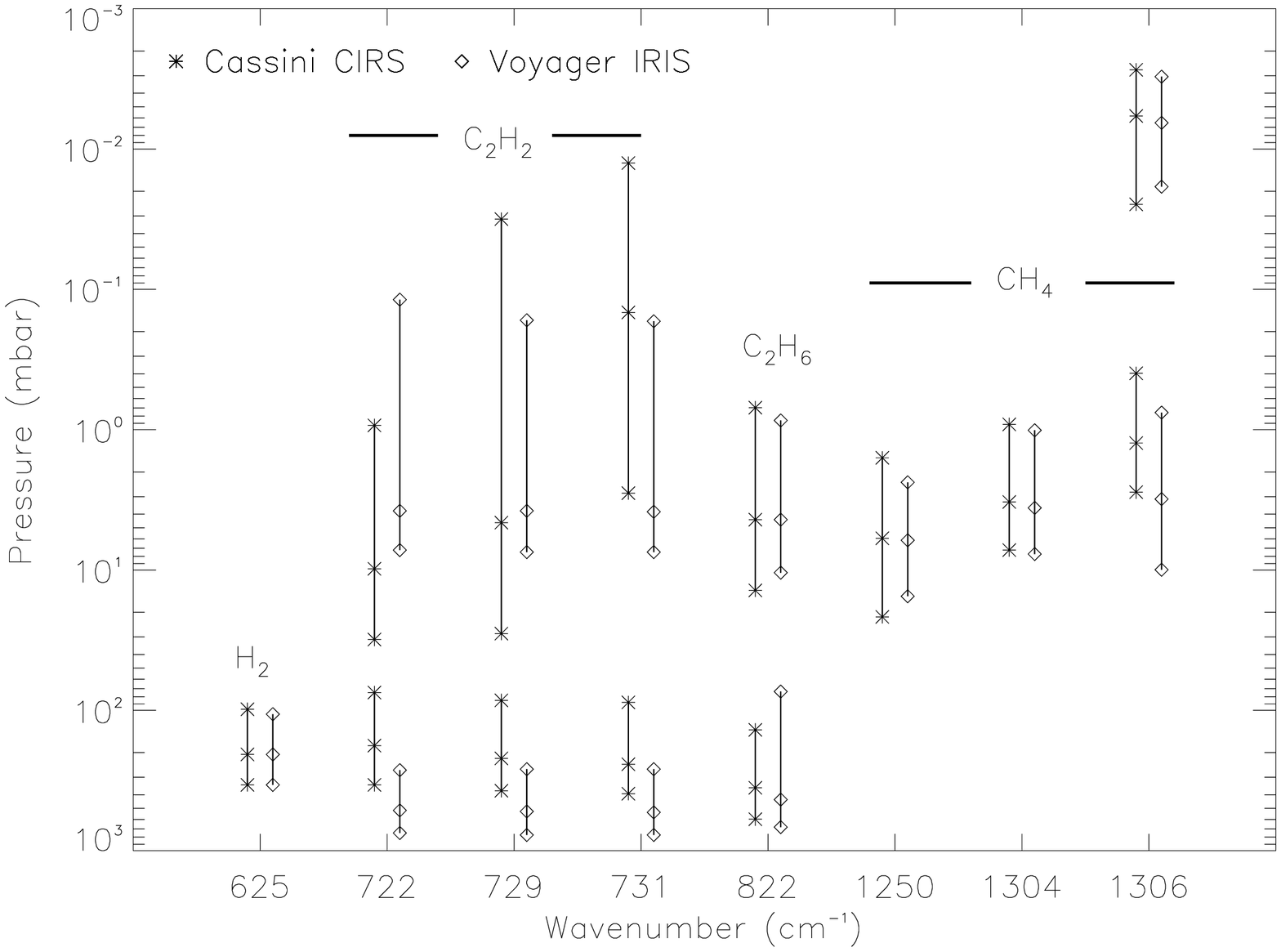, height=19cm, angle=90}
}
\end{figure}

Nixon et al. {\bf Figure \ref{fig:comp_fds_equat}}

\clearpage

\begin{figure}[t]\centerline{
\epsfig{file=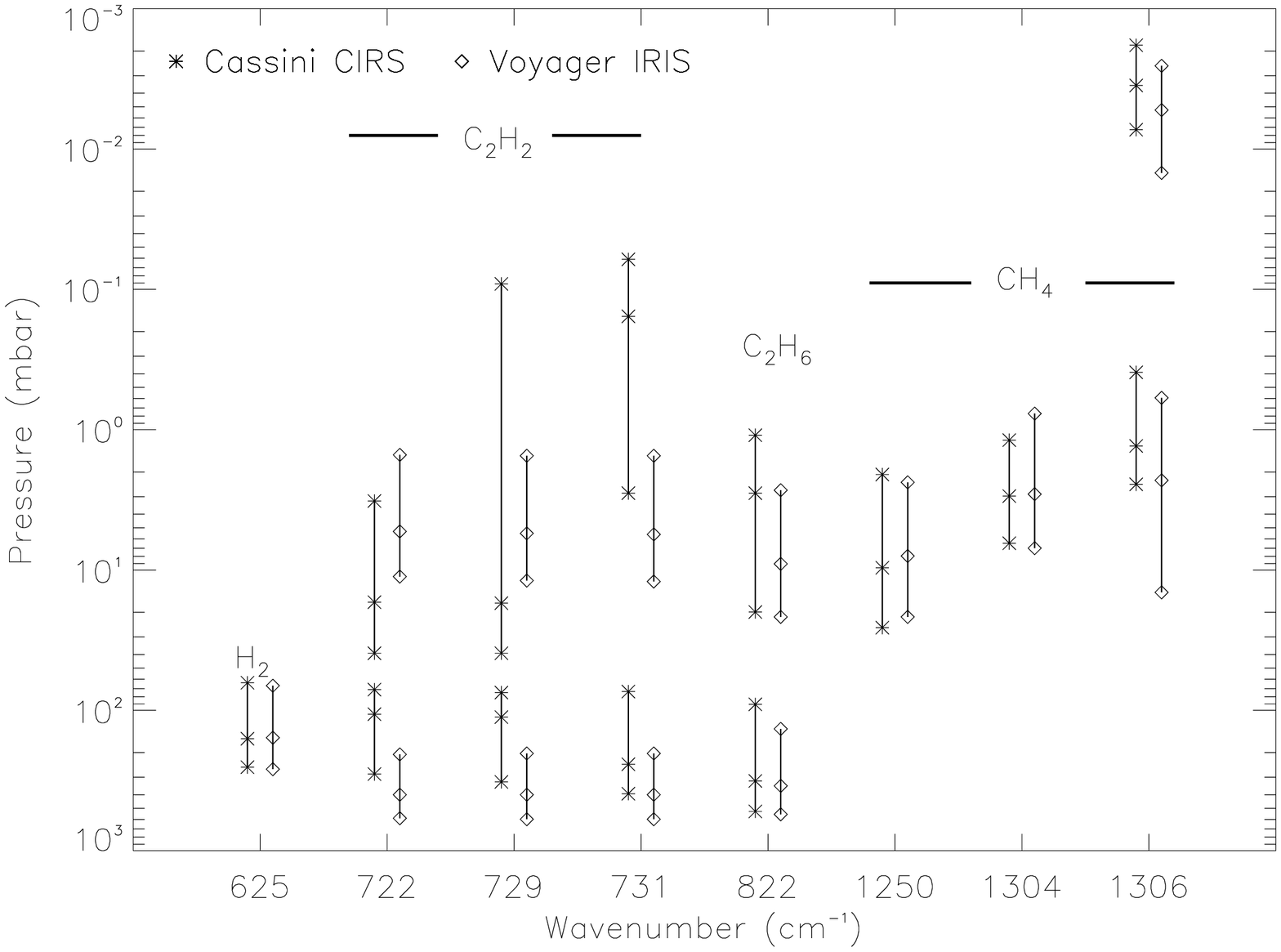, height=19cm, angle=90}
}
\end{figure}

Nixon et al. {\bf Figure \ref{fig:comp_fds_north}}

\clearpage

\begin{figure}[t]\centerline{
\epsfig{file=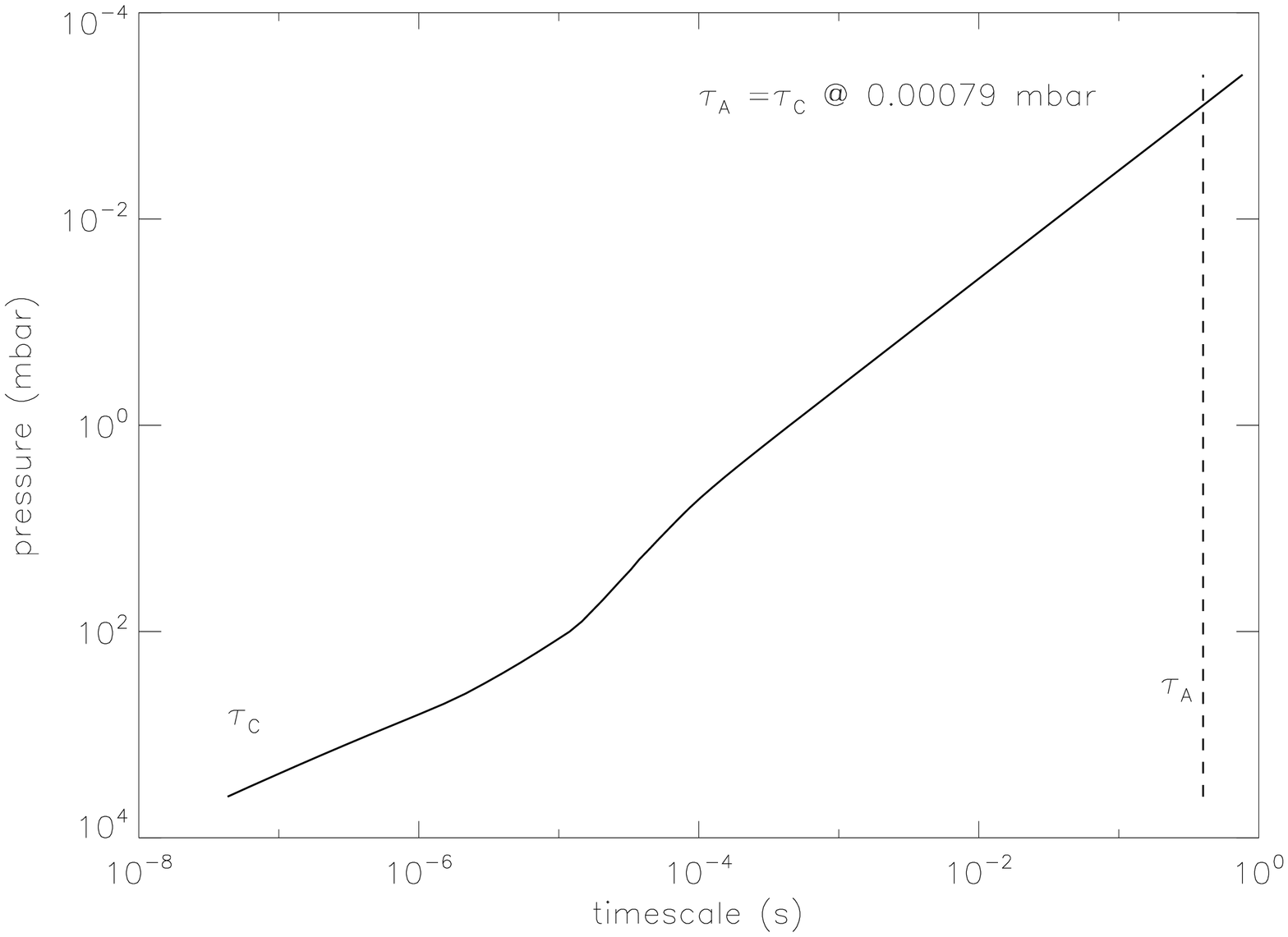, height=19cm, angle=90}
}
\end{figure}

Nixon et al. {\bf Figure \ref{fig:nlte}}

\clearpage

\begin{figure}[t]\centerline{
\epsfig{file=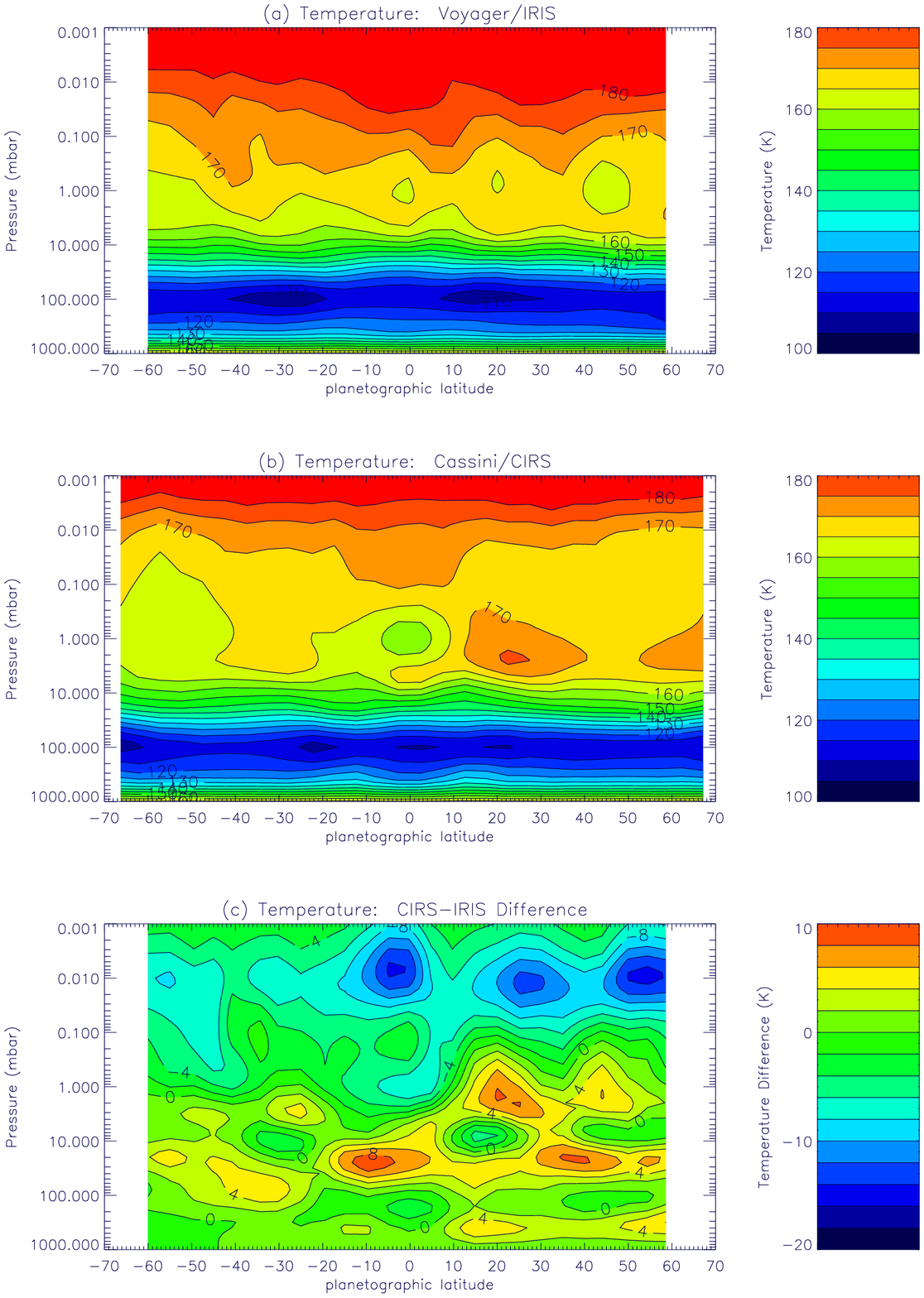, height=19cm, angle=0}
}
\end{figure}

Nixon et al. {\bf Figure \ref{fig:temp_contours}}

\clearpage

\begin{figure}[t]\centerline{
\epsfig{file=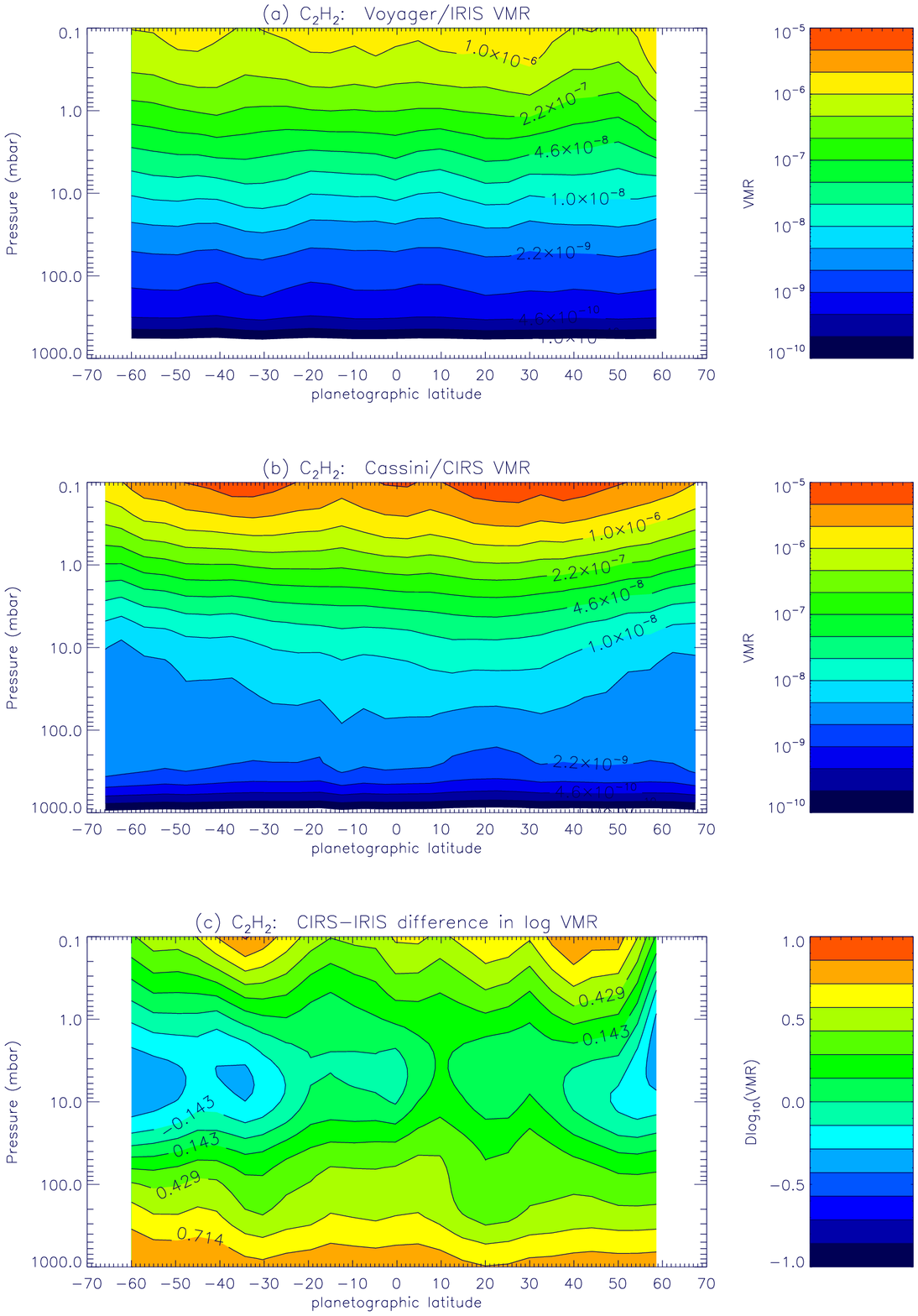, height=19cm, angle=0}
}
\end{figure}

Nixon et al. {\bf Figure \ref{fig:c2h2_contours}}

\clearpage

\begin{figure}[t]\centerline{
\epsfig{file=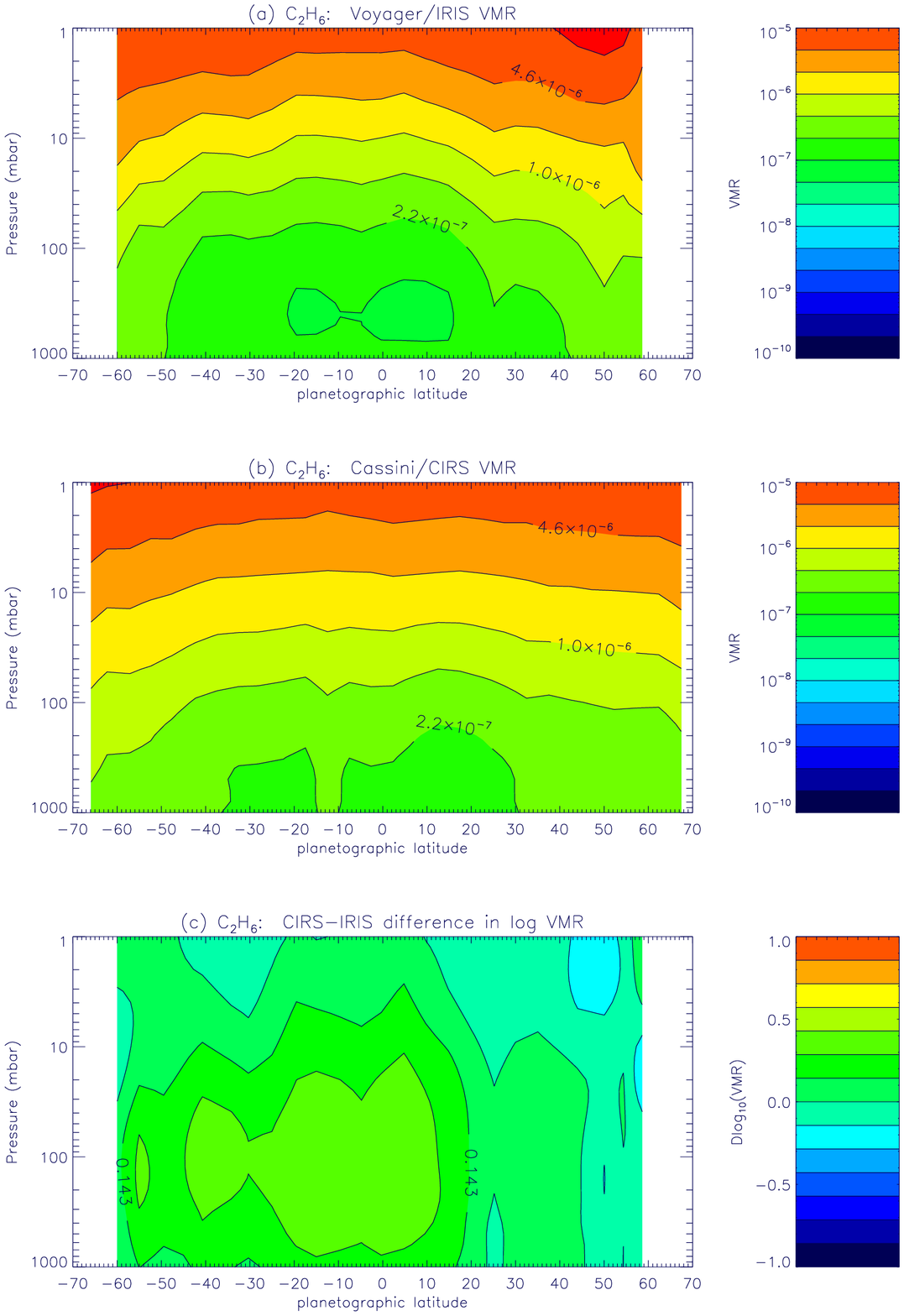, height=19cm, angle=0}
}
\end{figure}

Nixon et al. {\bf Figure \ref{fig:c2h6_contours}}

\clearpage

\begin{figure}[t]\centerline{
\epsfig{file=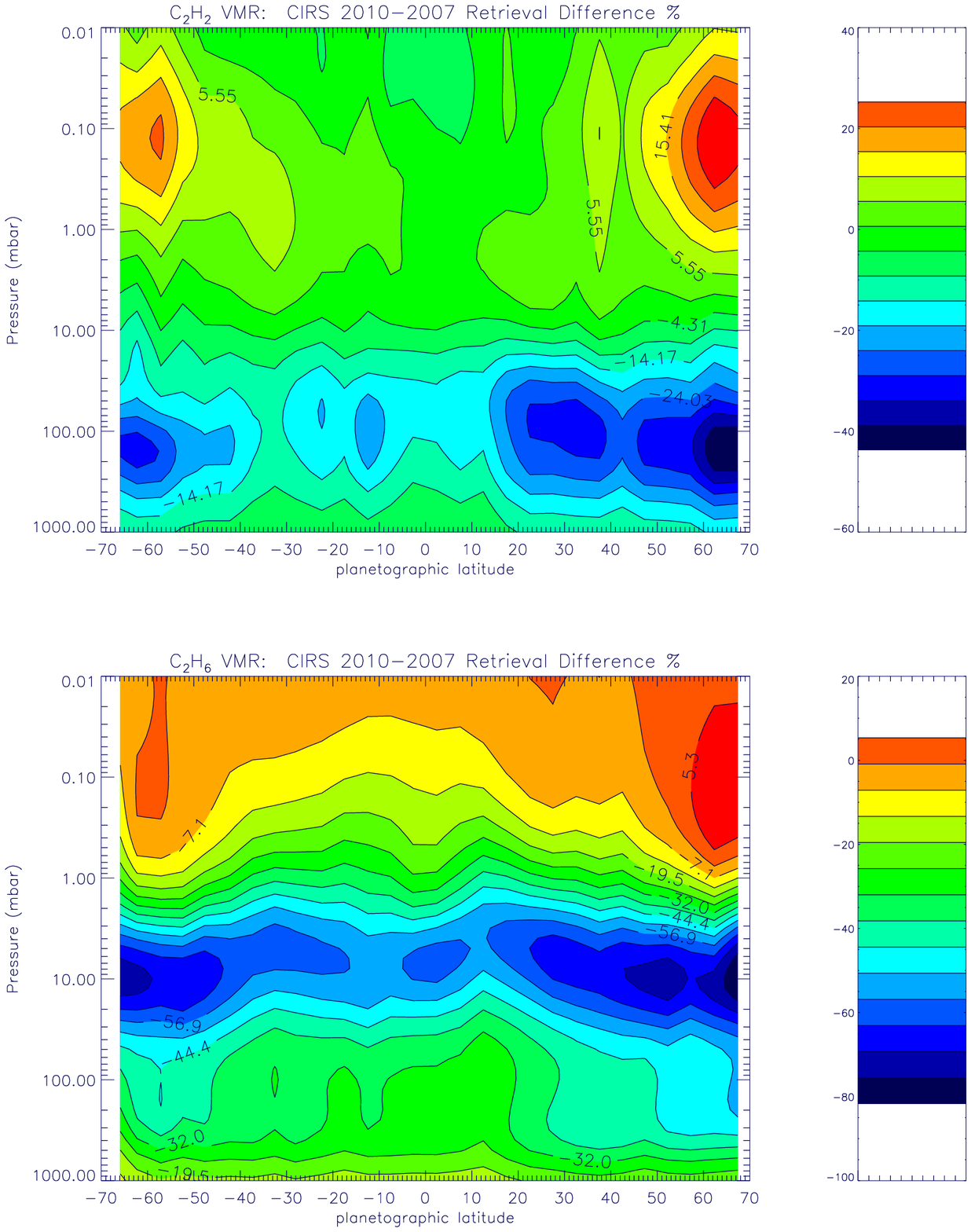, height=19cm, angle=0}
}
\end{figure}

Nixon et al. {\bf Figure \ref{fig:cirs_comp}}

\clearpage

\begin{figure}[t]\centerline{
\epsfig{file=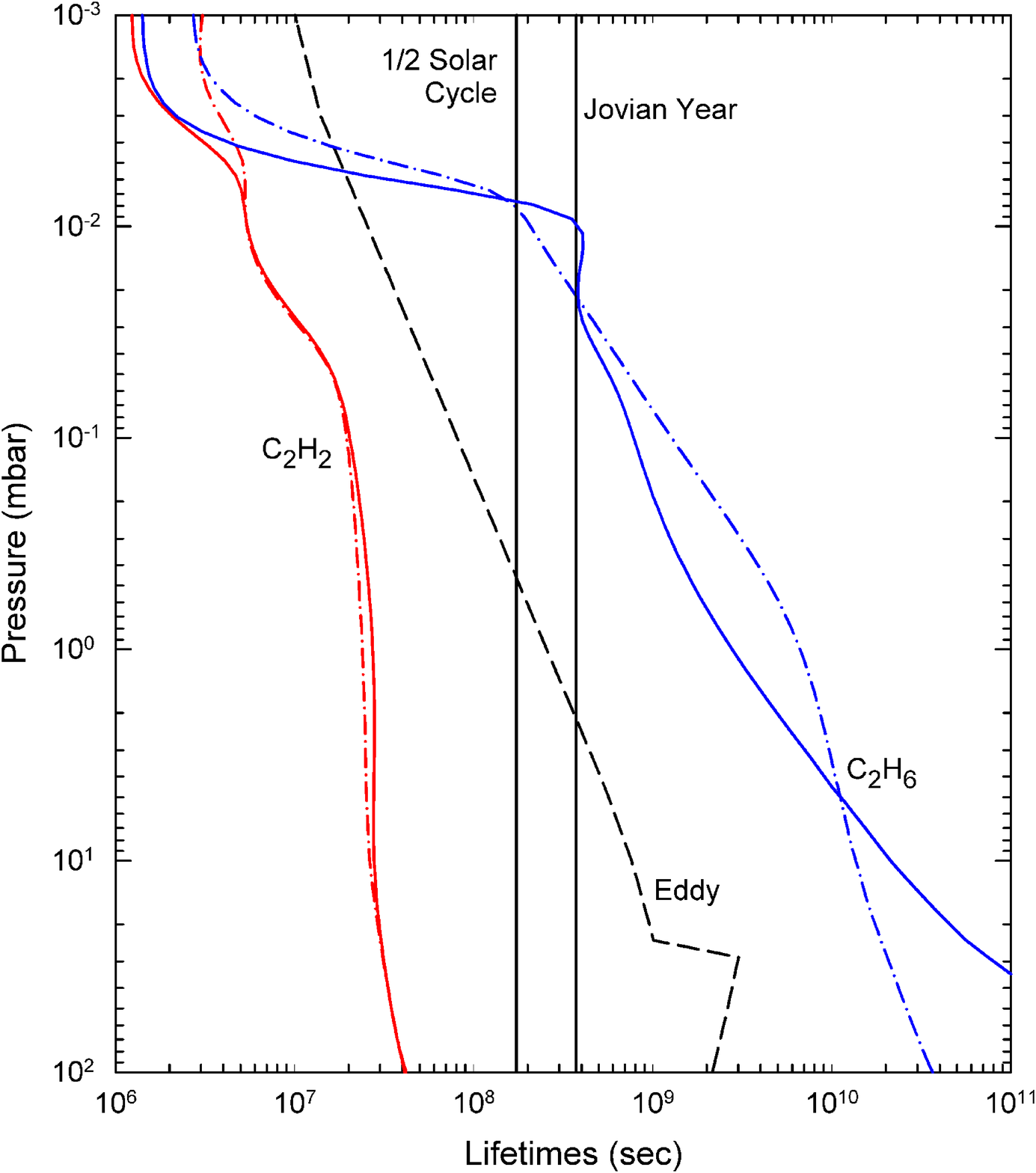, height=19cm, angle=0}
}
\end{figure}

Nixon et al. {\bf Figure \ref{fig:lifetimes}}

\end{document}